
%
%
%
%
%
%

\input amstex             

\def\Resetstrings{
    \def\present{ }\let\bgroup={\let\egroup=}
    \def\Astr{}\def\astr{}\def\Atest{}\def\atest{}%
    \def\Bstr{}\def\bstr{}\def\Btest{}\def\btest{}%
    \def\Cstr{}\def\cstr{}\def\Ctest{}\def\ctest{}%
    \def\Dstr{}\def\dstr{}\def\Dtest{}\def\dtest{}%
    \def\Estr{}\def\estr{}\def\Etest{}\def\etest{}%
    \def\Fstr{}\def\fstr{}\def\Ftest{}\def\ftest{}%
    \def\Gstr{}\def\gstr{}\def\Gtest{}\def\gtest{}%
    \def\Hstr{}\def\hstr{}\def\Htest{}\def\htest{}%
    \def\Istr{}\def\istr{}\def\Itest{}\def\itest{}%
    \def\Jstr{}\def\jstr{}\def\Jtest{}\def\jtest{}%
    \def\Kstr{}\def\kstr{}\def\Ktest{}\def\ktest{}%
    \def\Lstr{}\def\lstr{}\def\Ltest{}\def\ltest{}%
    \def\Mstr{}\def\mstr{}\def\Mtest{}\def\mtest{}%
    \def\Nstr{}\def\nstr{}\def\Ntest{}\def\ntest{}%
    \def\Ostr{}\def\ostr{}\def\Otest{}\def\otest{}%
    \def\Pstr{}\def\pstr{}\def\Ptest{}\def\ptest{}%
    \def\Qstr{}\def\qstr{}\def\Qtest{}\def\qtest{}%
    \def\Rstr{}\def\rstr{}\def\Rtest{}\def\rtest{}%
    \def\Sstr{}\def\sstr{}\def\Stest{}\def\stest{}%
    \def\Tstr{}\def\tstr{}\def\Ttest{}\def\ttest{}%
    \def\Ustr{}\def\ustr{}\def\Utest{}\def\utest{}%
    \def\Vstr{}\def\vstr{}\def\Vtest{}\def\vtest{}%
    \def\Wstr{}\def\wstr{}\def\Wtest{}\def\wtest{}%
    \def\Xstr{}\def\xstr{}\def\Xtest{}\def\xtest{}%
    \def\Ystr{}\def\ystr{}\def\Ytest{}\def\ytest{}%
}
\Resetstrings

\def\Refformat{
         \if\Jtest\present
             {\if\Vtest\present\journalarticleformat
                  \else\conferencereportformat\fi}
            \else\if\Btest\present\bookarticleformat
               \else\if\Rtest\present\technicalreportformat
                  \else\if\Itest\present\bookformat
                     \else\otherformat\fi\fi\fi\fi}

\def\Rpunct{
   \def\Lspace{ }%
   \def\Lperiod{ }
   \def\Lcomma{ }
   \def\Lquest{ }
   \def\Lcolon{ }
   \def\Lscolon{ }
   \def\Lbang{ }
   \def\Lquote{ }
   \def\Lqquote{ }
   \def\Lrquote{ }
   \def\Rspace{}%
   \def\Rperiod{.}
   \def\Rcomma{,}
   \def\Rquest{?}
   \def\Rcolon{:}
   \def\Rscolon{;}
   \def\Rbang{!}
   \def\Rquote{'}
   \def\Rqquote{"}
   \def\Rrquote{`}
   }

\def\Lpunct{
   \def\Lspace{}%
   \def\Lperiod{\unskip.}
   \def\Lcomma{\unskip,}
   \def\Lquest{\unskip?}
   \def\Lcolon{\unskip:}
   \def\Lscolon{\unskip;}
   \def\Lbang{\unskip!}
   \def\Lquote{\unskip'}
   \def\Lqquote{\unskip"}
   \def\Lrquote{\unskip`}
   \def\Rspace{\spacefactor=1000}%
   \def\Rperiod{\spacefactor=3000}
   \def\Rcomma{\spacefactor=1250}
   \def\Rquest{\spacefactor=3000}
   \def\Rcolon{\spacefactor=2000}
   \def\Rscolon{\spacefactor=1250}
   \def\Rbang{\spacefactor=3000}
   \def\Rquote{\spacefactor=1000}
   \def\Rqquote{\spacefactor=1000}
   \def\Rrquote{\spacefactor=1000}
   }

\def\Refstd{
     \def\Acomma{\unskip, }
     \def\Aand{\unskip\ and }
     \def\Aandd{\unskip\ and }
     \def\Ecomma{\unskip, }
     \def\Eand{\unskip\ and }
     \def\Eandd{\unskip\ and }
     \def\acomma{\unskip, }
     \def\aand{\unskip\ and }
     \def\aandd{\unskip\ and }
     \def\ecomma{\unskip, }
     \def\eand{\unskip\ and }
     \def\eandd{\unskip\ and }
     \def\Namecomma{\unskip, }
     \def\Nameand{\unskip\ and }
     \def\Nameandd{\unskip\ and }
     \def\Revcomma{\unskip, }
     \def\Initper{.\ }
     \def\Initgap{\dimen0=\spaceskip\divide\dimen0 by 2\hskip-\dimen0}%
   }

\def\Smallcapsaand{
     \def\Aand{\unskip\bgroup{\Smallcapsfont\ AND }\egroup}%
     \def\Aandd{\unskip\bgroup{\Smallcapsfont\ AND }\egroup}%
     \def\eand{\unskip\bgroup\Smallcapsfont\ AND \egroup}%
     \def\eandd{\unskip\bgroup\Smallcapsfont\ AND \egroup}%
   }

\def\Smallcapseand{
     \def\Eand{\unskip\bgroup\Smallcapsfont\ AND \egroup}%
     \def\Eandd{\unskip\bgroup\Smallcapsfont\ AND \egroup}%
     \def\aand{\unskip\bgroup\Smallcapsfont\ AND \egroup}%
     \def\aandd{\unskip\bgroup\Smallcapsfont\ AND \egroup}%
   }

   \def\Citefont{}
   \def\ACitefont{}
   \def\Authfont{}
   \def\Titlefont{}
   \def\Tomefont{\sl}
   \def\Volfont{}
   \def\Flagfont{}
   \def\Reffont{\rm}
   \def\Smallcapsfont{\sevenrm}
   \def\Flagstyle#1{\hangindent\parindent\indent\hbox to0pt
       {\hss[{\Flagfont#1}]\kern.5em}\ignorespaces}


\def\Citebrackets{\Rpunct
   \def\Lcitemark{\def\Cfont{\Citefont}[\bgroup\Cfont}
   \def\Rcitemark{\egroup]}
   \def\LAcitemark{\def\Cfont{\ACitefont}\bgroup\ACitefont}%
   \def\RAcitemark{\egroup}
   \def\LIcitemark{\egroup}
   \def\RIcitemark{\bgroup\Cfont}
   \def\Citehyphen{\egroup--\bgroup\Cfont}
   \def\Citecomma{\egroup,\hskip0pt\bgroup\Cfont}%
   \def\Citebreak{}
   }

\def\Citeparen{\Rpunct
   \def\Lcitemark{\def\Cfont{\Citefont}(\bgroup\Cfont}
   \def\Rcitemark{\egroup)}
   \def\LAcitemark{\def\Cfont{\ACitefont}\bgroup\ACitefont}%
   \def\RAcitemark{\egroup}
   \def\LIcitemark{\egroup}
   \def\RIcitemark{\bgroup\Cfont}
   \def\Citehyphen{\egroup--\bgroup\Cfont}
   \def\Citecomma{\egroup,\hskip0pt\bgroup\Cfont}%
   \def\Citebreak{}
   }

\def\Citesuper{\Lpunct
   \def\Lcitemark{\def\Cfont{\Citefont}\raise1ex\hbox\bgroup\bgroup\Cfont}%
   \def\Rcitemark{\egroup\egroup}
   \def\LAcitemark{\def\Cfont{\ACitefont}\bgroup\ACitefont}%
   \def\RAcitemark{\egroup}
   \def\LIcitemark{\egroup\egroup}
   \def\RIcitemark{\raise1ex\hbox\bgroup\bgroup\Cfont}%
   \def\Citehyphen{\egroup--\bgroup\Cfont}
   \def\Citecomma{\egroup,\hskip0pt\bgroup%
      \Cfont}
   \def\Citebreak{}
   }

\def\Citenamedate{\Rpunct
   \def\Lcitemark{
      \def\Citebreak{\egroup\ [\bgroup\Citefont}
      \def\Citecomma{\egroup]; 
         \bgroup\let\uchyph=1\Citefont}(\bgroup\let\uchyph=1\Citefont}%
   \def\Rcitemark{\egroup])}
   \def\LAcitemark{
      \def\Citebreak{\egroup\ [\bgroup\Citefont}\def\Citecomma{\egroup], %
         \bgroup\ACitefont }\bgroup\let\uchyph=1\ACitefont}%
   \def\RAcitemark{\egroup]}
  \def\Citehyphen{\egroup--\bgroup\Citefont}
   \def\LIcitemark{\egroup}
   \def\RIcitemark{\bgroup\Citefont}
   }


\Refstd\Rpunct
\def\Titlefont{\sl}\def\Volfont{\bf}\def\Tomefont{\Reffont}
   \def\Lcitemark{
      \let\uchyph=1\def\Cfont{\Citefont}%
      \def\Citebreak{\egroup\ [\bgroup\bf}%
      \def\Citecomma{\egroup]; \bgroup\Cfont }(\bgroup\Cfont}%
   \def\Rcitemark{\egroup])}
   \def\LAcitemark{
      \let\uchyph=1\def\Citebreak{\egroup\ [\bgroup\bf}%
      \def\Cfont{\ACitefont}%
      \def\Citecomma{\egroup], \bgroup\Cfont }\bgroup\Cfont}%
   \def\RAcitemark{\egroup]}
   \def\Citehyphen{\egroup--\bgroup\Cfont}
   \def\LIcitemark{\egroup}
   \def\RIcitemark{\bgroup\Cfont}

\def\journalarticleformat{\Reffont\let\uchyph=1\parindent=1.25pc\def\Comma{}%

\sfcode`\.=1000\sfcode`\?=1000\sfcode`\!=1000\sfcode`\:=1000\sfcode`\;=1000\sfcode`\,=1000
                \par\vfil\penalty-200\vfilneg
      \if\Atest\present\bgroup\Authfont\Astr\egroup\def\Comma{\unskip, }\fi%
       \if\Dtest\present\unskip\hskip.16667em[\bgroup\bf\Dstr\ifcat\Ftrail
a\Ftrail\fi\egroup]\def\Comma{, }\fi%
        \if\Ttest\present\Comma\bgroup\Titlefont\Tstr\egroup\def\Comma{, }\fi%

\if\etest\present\if\Ttest\present{}\fi\hskip.16667em(\bgroup\estr\egroup)\def\Comma{\unskip, }\fi%
          \if\Jtest\present\Comma\bgroup\Tomefont\Jstr\/\egroup\def\Comma{,
}\fi%

\if\Vtest\present\if\Jtest\present\hskip.2em\else\Comma\fi\bgroup\Volfont\Vstr\egroup\def\Comma{, }\fi%
            \if\Ptest\present\bgroup, \Pstr\egroup\def\Comma{, }\fi%
             \if\ttest\present\Comma\bgroup\Titlefont\tstr\egroup\def\Comma{,
}\fi%
              \if\jtest\present\Comma\bgroup\Tomefont\jstr\/\egroup\def\Comma{,
}\fi%

\if\vtest\present\if\jtest\present\hskip.2em\else\Comma\fi\bgroup\Volfont\vstr\egroup\def\Comma{, }\fi%

\if\dtest\present\hskip.16667em(\bgroup\dstr\egroup)\def\Comma{, }\fi%
                 \if\ptest\present\bgroup, \pstr\egroup\def\Comma{, }\fi%
                  \if\Gtest\present{\Comma Gov't ordering no.
}\bgroup\Gstr\egroup\def\Comma{, }\fi%
                   \if\Mtest\present\Comma MR \#\bgroup\Mstr\egroup\def\Comma{,
}\fi%
                    \if\Otest\present{\Comma\bgroup\Ostr\egroup.}\else{.}\fi%
                     \vskip3ptplus1ptminus1pt}

\def\conferencereportformat{\Reffont\let\uchyph=1\parindent=1.25pc\def\Comma{}%

\sfcode`\.=1000\sfcode`\?=1000\sfcode`\!=1000\sfcode`\:=1000\sfcode`\;=1000\sfcode`\,=1000
                \par\vfil\penalty-200\vfilneg
      \if\Atest\present\bgroup\Authfont\Astr\egroup\def\Comma{\unskip, }\fi%
       \if\Dtest\present\unskip\hskip.16667em[\bgroup\bf\Dstr\ifcat
a\Ftrail\Ftrail\fi\egroup]\def\Comma{, }\fi%
        \if\Ttest\present\Comma\bgroup\Titlefont\Tstr\egroup\def\Comma{, }\fi%
         \if\Jtest\present\Comma\bgroup\Tomefont\Jstr\/\egroup\def\Comma{,
}\fi%
          \if\Ctest\present\Comma\bgroup\Cstr\egroup\def\Comma{, }\fi%
           \if\Mtest\present\Comma MR \#\bgroup\Mstr\egroup\def\Comma{, }\fi%
            \if\Otest\present{\Comma\bgroup\Ostr\egroup.}\else{.}\fi%
             \vskip3ptplus1ptminus1pt}

\def\bookarticleformat{\Reffont\let\uchyph=1\parindent=1.25pc\def\Comma{}%

\sfcode`\.=1000\sfcode`\?=1000\sfcode`\!=1000\sfcode`\:=1000\sfcode`\;=1000\sfcode`\,=1000
                \par\vfil\penalty-200\vfilneg
      \if\Atest\present\bgroup\Authfont\Astr\egroup\def\Comma{\unskip, }\fi%
       \if\Dtest\present\unskip\hskip.16667em[\bgroup\bf\Dstr\ifcat
a\Ftrail\Ftrail\fi\egroup]\def\Comma{, }\fi%
        \if\Ttest\present\Comma\bgroup\Titlefont\Tstr\egroup\def\Comma{, }\fi%

\if\etest\present\if\Ttest\present\fi\hskip.2em(\bgroup\estr\egroup)\def\Comma{\unskip, }\fi%
          \if\Btest\present\Comma in
``\bgroup\Tomefont\Bstr\/\egroup''\def\Comma{\unskip, }\fi%
           \if\otest\present\ \bgroup\ostr\egroup\def\Comma{, }\fi%
            \if\Etest\present\Comma\bgroup\Estr\egroup\unskip,
\ifnum\Ecnt>1eds.\else ed.\fi\def\Comma{, }\fi%
             \if\Stest\present\Comma\bgroup\Sstr\egroup\def\Comma{, }\fi%
              \if\Vtest\present\Comma vol.
\bgroup\Volfont\Vstr\egroup\def\Comma{, }\fi%
               \if\Ntest\present\Comma no.
\bgroup\Volfont\Nstr\egroup\def\Comma{, }\fi%
                \if\Itest\present\Comma\bgroup\Istr\egroup\def\Comma{, }\fi%
                 \if\Ctest\present\Comma\bgroup\Cstr\egroup\def\Comma{, }\fi%
                  \if\Ptest\present\Comma\Pstr\def\Comma{, }\fi%

\if\ttest\present\Comma\bgroup\Titlefont\Tstr\egroup\def\Comma{, }\fi%
                    \if\btest\present\Comma in
``\bgroup\Tomefont\bstr\egroup''\def\Comma{, }\fi%
                     \if\atest\present\Comma\bgroup\astr\egroup\unskip,
\if\acnt\present eds.\else ed.\fi\def\Comma{, }\fi%
                      \if\stest\present\Comma\bgroup\sstr\egroup\def\Comma{,
}\fi%
                       \if\vtest\present\Comma vol.
\bgroup\Volfont\vstr\egroup\def\Comma{, }\fi%
                        \if\ntest\present\Comma no.
\bgroup\Volfont\nstr\egroup\def\Comma{, }\fi%
                         \if\itest\present\Comma\bgroup\istr\egroup\def\Comma{,
}\fi%

\if\ctest\present\Comma\bgroup\cstr\egroup\def\Comma{, }\fi%

\if\dtest\present\Comma\bgroup\dstr\egroup\def\Comma{, }\fi%
                            \if\ptest\present\Comma\pstr\def\Comma{, }\fi%
                             \if\Gtest\present{\Comma Gov't ordering no.
}\bgroup\Gstr\egroup\def\Comma{, }\fi%
                              \if\Mtest\present\Comma MR
\#\bgroup\Mstr\egroup\def\Comma{, }\fi%

\if\Otest\present{\Comma\bgroup\Ostr\egroup.}\else{.}\fi%
                                \vskip3ptplus1ptminus1pt}

\def\bookformat{\Reffont\let\uchyph=1\parindent=1.25pc\def\Comma{}%

\sfcode`\.=1000\sfcode`\?=1000\sfcode`\!=1000\sfcode`\:=1000\sfcode`\;=1000\sfcode`\,=1000
                \par\vfil\penalty-200\vfilneg
       \if\Atest\present\bgroup\Authfont\Astr\egroup\def\Comma{\unskip, }%

\else\if\Etest\present\bgroup\def\Eand{\Aand}\def\Eandd{\Aandd}\Authfont\Estr\egroup\unskip, \ifnum\Ecnt>1eds.\else ed.\fi\def\Comma{, }%
                 \else\if\Itest\present\bgroup\Authfont\Istr\egroup\def\Comma{,
}\fi\fi\fi%
         \if\Dtest\present\unskip\hskip.16667em[\bgroup\bf\Dstr\ifcat
a\Ftrail\Ftrail\fi\egroup]\def\Comma{, }\fi%
          \if\Ttest\present\Comma\bgroup\Titlefont\Tstr\egroup\def\Comma{, }%

\else\if\Btest\present\Comma\bgroup\Titlefont\Bstr\/\egroup\def\Comma{\unskip,
}\fi\fi%
            \if\otest\present\ \bgroup\ostr\egroup\def\Comma{, }\fi%

\if\etest\present\hskip.2em(\bgroup\estr\egroup)\def\Comma{\unskip, }\fi%
              \if\Stest\present\Comma\bgroup\Sstr\egroup\def\Comma{, }\fi%
               \if\Vtest\present\Comma vol.
\bgroup\Volfont\Vstr\egroup\def\Comma{, }\fi%
                \if\Ntest\present\Comma no.
\bgroup\Volfont\Nstr\egroup\def\Comma{, }\fi%
                 \if\Atest\present\if\Itest\present
                         \Comma\bgroup\Istr\egroup\def\Comma{\unskip, }\fi%
                      \else\if\Etest\present\if\Itest\present
                              \Comma\bgroup\Istr\egroup\def\Comma{\unskip,
}\fi\fi\fi%
                     \if\Ctest\present\Comma\bgroup\Cstr\egroup\def\Comma{,
}\fi%

\if\ttest\present\Comma\bgroup\Titlefont\tstr\egroup\def\Comma{, }%

\else\if\btest\present\Comma\bgroup\Titlefont\bstr\egroup\def\Comma{, }\fi\fi%
                         \if\stest\present\Comma\bgroup\sstr\egroup\def\Comma{,
}\fi%
                          \if\vtest\present\Comma vol.
\bgroup\Volfont\vstr\egroup\def\Comma{, }\fi%
                           \if\ntest\present\Comma no.
\bgroup\Volfont\nstr\egroup\def\Comma{, }\fi%

\if\itest\present\Comma\bgroup\istr\egroup\def\Comma{, }\fi%

\if\ctest\present\Comma\bgroup\cstr\egroup\def\Comma{, }\fi%

\if\dtest\present\Comma\bgroup\dstr\egroup\def\Comma{, }\fi%
                               \if\Gtest\present{\Comma Gov't ordering no.
}\bgroup\Gstr\egroup\def\Comma{, }\fi%
                                \if\Mtest\present\Comma MR
\#\bgroup\Mstr\egroup\def\Comma{, }\fi%

\if\Otest\present{\Comma\bgroup\Ostr\egroup.}\else{.}\fi%
                                  \vskip3ptplus1ptminus1pt}

\def\technicalreportformat{\Reffont\let\uchyph=1\parindent=1.25pc\def\Comma{}%

\sfcode`\.=1000\sfcode`\?=1000\sfcode`\!=1000\sfcode`\:=1000\sfcode`\;=1000\sfcode`\,=1000
                \par\vfil\penalty-200\vfilneg
      \if\Atest\present\bgroup\Authfont\Astr\egroup\def\Comma{\unskip, }%

\else\if\Etest\present\bgroup\def\Eand{\Aand}\def\Eandd{\Aandd}\Authfont\Estr\egroup\unskip, \ifnum\Ecnt>1eds.\else ed.\fi\def\Comma{, }%
                 \else\if\Itest\present\bgroup\Authfont\Istr\egroup\def\Comma{,
}\fi\fi\fi%
         \if\Dtest\present\unskip\hskip.16667em[\bgroup\bf\Dstr\ifcat
a\Ftrail\Ftrail\fi\egroup]\def\Comma{, }\fi%
          \if\Ttest\present\Comma\bgroup\Titlefont\Tstr\egroup\def\Comma{,
}\fi%
           \if\Atest\present\if\Itest\present
                   \Comma\bgroup\Istr\egroup\def\Comma{, }\fi%
                \else\if\Etest\present\if\Itest\present
                        \Comma\bgroup\Istr\egroup\def\Comma{, }\fi\fi\fi%
            \if\Rtest\present\Comma\bgroup\Rstr\egroup\def\Comma{, }\fi%
             \if\Ctest\present\Comma\bgroup\Cstr\egroup\def\Comma{, }\fi%
              \if\ttest\present\Comma\bgroup\Titlefont\tstr\egroup\def\Comma{,
}\fi%
               \if\itest\present\Comma\bgroup\istr\egroup\def\Comma{, }\fi%
                \if\rtest\present\Comma\bgroup\rstr\egroup\def\Comma{, }\fi%
                 \if\ctest\present\Comma\bgroup\cstr\egroup\def\Comma{, }\fi%
                  \if\dtest\present\Comma\bgroup\dstr\egroup\def\Comma{, }\fi%
                   \if\Gtest\present{\Comma Gov't ordering no.
}\bgroup\Gstr\egroup\def\Comma{, }\fi%
                    \if\Mtest\present\Comma MR
\#\bgroup\Mstr\egroup\def\Comma{, }\fi%
                     \if\Otest\present{\Comma\bgroup\Ostr\egroup.}\else{.}\fi%
                      \vskip3ptplus1ptminus1pt}

\def\otherformat{\Reffont\let\uchyph=1\parindent=1.25pc\def\Comma{}%

\sfcode`\.=1000\sfcode`\?=1000\sfcode`\!=1000\sfcode`\:=1000\sfcode`\;=1000\sfcode`\,=1000
                \par\vfil\penalty-200\vfilneg
      \if\Atest\present\bgroup\Authfont\Astr\egroup\def\Comma{\unskip, }%

\else\if\Etest\present\bgroup\def\Eand{\Aand}\def\Eandd{\Aandd}\Authfont\Estr\egroup\unskip, \ifnum\Ecnt>1eds.\else ed.\fi\def\Comma{, }%
                 \else\if\Itest\present\bgroup\authfont\Istr\egroup\def\Comma{,
}\fi\fi\fi%
         \if\Dtest\present\unskip\hskip.16667em[\bgroup\bf\Dstr\ifcat
a\Ftrail\Ftrail\fi\egroup]\def\Comma{, }\fi%
          \if\Ttest\present\Comma\bgroup\Titlefont\Tstr\egroup\def\Comma{,
}\fi%
           \if\Atest\present\if\Itest\present
                   \Comma\bgroup\Istr\egroup\def\Comma{, }\fi%
                \else\if\Etest\present\if\Itest\present
                        \Comma\bgroup\Istr\egroup\def\Comma{, }\fi\fi\fi%
                \if\Ctest\present\Comma\bgroup\Cstr\egroup\def\Comma{, }\fi%
                 \if\Gtest\present{\Comma Gov't ordering no.
}\bgroup\Gstr\egroup\def\Comma{, }\fi%
                  \if\Mtest\present\Comma MR \#\bgroup\Mstr\egroup\def\Comma{,
}\fi%
                   \if\Otest\present{\Comma\bgroup\Ostr\egroup.}\else{.}\fi%
                    \vskip3ptplus1ptminus1pt}

\magnification = 1200
\NoBlackBoxes

\def\today{24 June 1994}

\newcount\tagc \global\tagc=0
\def\tagno{\global\advance\tagc by 1 \tag\number\tagc}
\newcount\headc \headc=0
\def\headno{\global\advance\headc by 1 \number\headc . \global\propc=0}
\newcount\propc \propc=0
\def\propno{\global\advance\propc by 1 \number\headc .\number\propc}
\newcount\figc \figc=1
\def\figno{\botcaption{Figure \number\figc}
\endcaption \global\advance\figc by 1}

\def\nextfigure#1{\edef#1{\number\figc}}
\def\name#1{\xdef#1{{\number\headc .\number\propc}}}

\def\End{\operatorname{End}}
\def\Hom{\operatorname{Hom}}

\def\tr{\operatorname{tr}}

\def\C{{\Bbb C}}

\def\Z{{\Bbb Z}}

\define\ring{{\Bbb F}}

\define\spacehat{{\sphat\;}}
\define\placemarker{-}
\define\SL{{\hbox{$\frak sl$}}}
\accentedsymbol\ahathat{\Hat{\Hat a}}
\accentedsymbol\bhathat{\Hat{\Hat b}}

\input epsf.tex

\documentstyle{amsppt}

\topmatter
\title Invariants of\\
piecewise-linear 3-manifolds \endtitle
\author John W. Barrett and Bruce W. Westbury \endauthor
\date\today\enddate
\address
Department of Mathematics,
University of Nottingham,
University Park,
Nottingham,
NG7 2RD
\endaddress
\email jwb\@maths.nott.ac.uk \endemail
\address
Department of Mathematics,
University of Nottingham,
University Park,
Nottingham,
NG7 2RD
\endaddress
\email bww\@maths.nott.ac.uk \endemail
\toc
\head {0.} Introduction \endhead
\head {1.} State sum models \endhead
\head {2.} Spherical categories \endhead
\head {3.} Symmetries of simplicial invariants\endhead
\head {4.} Piecewise-linear manifolds \endhead
\head {5.} Invariants of manifolds \endhead
\head {6.} Spherical Hopf algebras \endhead
\endtoc
\endtopmatter

\document
\head {0.} Introduction \endhead

The purpose of this paper is to present an algebraic framework for
constructing invariants of closed oriented 3-manifolds. The construction
is in the spirit of topological field theory and the invariant is
calculated from a triangulation of the 3-manifold. The data for the
construction of the invariant is a tensor category with a condition on
the duals, which we have called a spherical category. The definition of
a spherical category and a coherence theorem needed in this paper are given
in \Lcitemark Barrett\Nameand Westbury\Citebreak 1993\Rcitemark \Rspace{}.

There are two classes of examples of spherical categories
discussed in this paper. The first examples are given
by the quantised enveloping algebra of a semisimple Lie algebra, and the
second are given by an involutive Hopf algebra. In the first case,
the invariant for $\SL_2$ defined in this paper is the Turaev-Viro
invariant
\Lspace \Lcitemark Turaev\Nameand Viro\Citebreak 1992\Rcitemark \Rspace{}.
 This invariant is known to distinguish lens spaces of
the same homotopy type which already shows that the invariants in
this paper are not trivial. The problem of generalising the Turaev-Viro
invariant to other quantised enveloping algebras has also been considered
by\Lspace \Lcitemark Durhuus\Namecomma Jakobsen\Nameandd Nest\Citebreak
1993\Rcitemark \Rspace{} and
\Lcitemark  Yetter\Citebreak 1993\Rcitemark \Rspace{}.

A noteworthy feature of our construction is that it does not require a
braiding; the notion of a spherical category is more general than the
notion of a ribbon category. A simple example of this is the category
of representations of the convolution algebra of a non-abelian finite
group. This is a spherical category but does not admit a braiding because
the representation ring of the algebra is not commutative.

The second class of examples gives a manifold
invariant for any involutive Hopf algebra whose dimension is non-zero and
finite. It is shown in
\Lcitemark Barrett\Nameand Westbury\Citebreak 1994\Rcitemark \Rspace{}
 that this invariant is essentially the same
as the invariant defined in
\Lspace \Lcitemark Kuperberg\Citebreak 1991\Rcitemark \Rspace{}.

Another feature of this paper is that we prove invariance from a finite
list of moves on triangulations. The idea of working directly with
triangulations dates from the earlier work of\Lspace \Lcitemark
Ponzano\Nameand Regge\Citebreak 1968\Rcitemark \Rspace{} and
\Lcitemark Moussouris\Citebreak 1983\Rcitemark \Rspace{}
on the recoupling theory of Lie groups.
 The moves on triangulations replace the
Matveev moves on special spines which are used in
\Lspace \Lcitemark Turaev\Nameand Viro\Citebreak 1992\Rcitemark \Rspace{}
and\Lspace \Lcitemark Yetter\Citebreak 1993\Rcitemark \Rspace{}.
This approach shows that the invariants are defined for 3-manifolds which
may have point singularities of a prescribed type.

Finally, it is worth noting the two cases for which there is a known
relationship with the invariants which have been defined using surgery
presentations of a framed manifold and invariants of links. It is shown in
\Lspace \Lcitemark Walker\Citebreak 1990\Rcitemark \Rspace{},
\Lcitemark Turaev\Citebreak 1992\Rcitemark \Rspace{},
\Lspace \Lcitemark Roberts\Citebreak 1993\Rcitemark \Rspace{}
that the Turaev-Viro invariant is the square of the modulus of the
Witten invariant for $\SL_2$ which was defined in
\Lspace \Lcitemark Reshetikhin\Nameand Turaev\Citebreak 1991\Rcitemark
\Rspace{}. Also, direct
calculations show that the surgery invariant for the quantum double
of the group algebra of a finite group is equal to the invariant defined
in this paper for the group algebra itself.

\head\headno State Sum Models \endhead
In this paper we have assumed that $\ring$ is a field
and that a vector space is a finite dimensional vector space over $\ring$.

\definition{Definition \propno} A complex is a finite set of elements called
vertices, together with a subset of the set of all subsets. These are called
simplices. This is required to have the property that any subset of a simplex
is a simplex. \enddefinition

\definition{Definition \propno} A simplicial complex is a complex together
with a total ordering on the vertices of each simplex such that the ordering
on the vertices on any face of a simplex is the ordering induced from the
ordering on the vertices of the simplex. \enddefinition

Let $\sigma$ be an $n$-simplex in a simplicial complex. Then for $0\leqslant
i\leqslant n$ define $\partial_i \sigma$ to be the face obtained by omitting
the $i$-th vertex. These satisfy
 $$\partial_i\partial_j\sigma= \partial_{j-
1}\partial_i\sigma \qquad\text{if $i<j$}.$$

A simplicial complex is an example of the more general notion of simplicial
set. This explains the use of the adjective `simplicial' for the notion of a
complex with an ordering.
In the following, the adjective `combinatorial' will be used to refer to
complexes, reserving `simplicial' for simplicial
complexes. Combinatorial maps are maps of complexes and simplicial maps are
maps of simplicial complexes, i.e., combinatorial maps which preserve
orderings.
 For example, a single simplex considered as a simplicial complex has no
symmetries, whereas the corresponding complex admits the permutations as its
symmetries.

All manifolds are compact, oriented, piecewise-linear manifolds of
dimension three, unless stated otherwise.
For background on piecewise-linear
manifolds we refer the reader to\Lspace \Lcitemark Rourke\Nameand
Sanderson\Citebreak 1982\Rcitemark \Rspace{}. In line with the
terminology explained above, a simplicial manifold is a simplicial
complex whose geometric realisation is a piecewise-linear manifold, together
with an orientation.

Our notation for the orientation of a simplex is fixed as follows. The
standard $(n+1)$-simplex $(012\ldots n)$ with vertices $\{0,1,2,\ldots n\}$
has a standard orientation $(+)$. The opposite orientation is indicated
with a minus
$(-)$. The standard (oriented) tetrahedron $+(0123)$ has boundary
$$(123) - (023) + (013) - (012).$$
The signs indicate the induced orientation of the boundary of $+(0123)$.
The tetrahedron $-(0123)$ has the opposite orientation for the boundary,
$$- (123) + (023) - (013) + (012).$$
In an oriented closed manifold, each triangle is in the boundary of
exactly two tetrahedra, with each sign $+$ or $-$ occuring once.

The data for a state sum model consists of three parts, a set of labels $I$,
a set of state spaces for a triangle, and a set of partition functions for a
tetrahedron.

\definition{Definition \propno}
A labelled simplicial complex is a simplicial complex together with a function
which assigns an element of $I$ to each edge.
\enddefinition

\definition{Definition \propno} Let $T(a,b,c)$ be the standard oriented
triangle $+(012)$ labelled by
$\partial_0T\mapsto a,\partial_1T\mapsto b,\partial_2T\mapsto c$.
The state space for this labelled triangle is a vector space, $H(a,b,c)$.
The state space for the oppositely oriented triangle $-T(a,b,c)$ is
defined to be the dual vector space, $H^*(a,b,c)$. \enddefinition

\definition{Definition \propno} Let $A$ be the standard oriented tetrahedron
$+(0123)$ with the edge $\partial_i\partial_j\sigma$ labelled by $e_{ij}$.
The partition function of this labelled tetrahedron is defined to be
a linear map
$$\multline\left\{\matrix e_{01}&e_{02}&e_{12}\\e_{23}&e_{13}&e_{03}
\endmatrix\right\}_+ \colon \\
H(e_{23},e_{03},e_{02})
\otimes H(e_{12},e_{02},e_{01})
\to
H(e_{23},e_{13},e_{12} )
\otimes H(e_{13},e_{03},e_{01}).
\endmultline$$

The partition function of $-A$ with the same labelling is defined to be
a linear map
$$\multline\left\{\matrix e_{01}&e_{02}&e_{12}\\e_{23}&e_{13}&e_{03}
\endmatrix\right\}_- \colon \\
H(e_{23},e_{13},e_{12} )
\otimes H(e_{13},e_{03},e_{01})
\to
H(e_{23},e_{03},e_{02})
\otimes H(e_{12},e_{02},e_{01}).
\endmultline$$
\enddefinition

In the definition, the four factors
in the tensor products correspond to each of the four faces. Also, the two
factors in the domain of the linear map correspond to the two faces with sign
$-$ in the boundary of the tetrahedron, and the two factors in the range
correspond to the two faces with sign $+$.

\definition{Definition \propno\name\simplicialinvariant}
The data for a state sum model determines an element $Z(M)\in\ring$ for each
labelled simplicial closed manifold $M$. This is called the simplicial
invariant of the labelled manifold.
Let $V(M)$ be the tensor product over the set of triangles of
$M$ of the state space for each triangle. For each tetrahedron in $M$, take
the partition function of the labelled standard tetrahedron, $A$ or $-A$,
to which it is isomorphic.

The tensor product over this set of partition
functions is a linear map $V(M)\to V^\pi(M)$, where $V^\pi(M)$ is defined
in the same way as $V(M)$ but with the factors permuted by some permutation
$\pi$. This uses the fact that in a closed oriented manifold,
each triangle is in the boundary of two tetrahedra, each with
opposite orientation.
There is a unique standard linear map $V^\pi(M)\to V(M)$
given by iterating the standard twist $P\colon x\otimes y\mapsto y\otimes x$.
This defines a linear map $V(M)\to V(M)$ and the element $Z(M)$ is defined
to be the trace of this linear map.
\enddefinition

Note that if $A\colon X\to Y$ and $B\colon Y\to X$ are linear maps, then
$$\tr^{X\otimes Y}(P(A\otimes B))=\tr^X(AB)=\tr^Y(BA).$$

This also introduces the notation used throughout the paper that the
map $X$ composed with map $Y$ is written $XY$ (not $Y\circ X$).

A state sum invariant of a closed manifold is obtained by a weighted sum of
these elements $Z(M)$ over a class of labellings. This state sum invariant
is defined in section 5.

\head\headno Spherical categories \endhead

The data which defines the
state sum model is a spherical category, whose definition is obtained
by axiomatising the properties of the category of representations of the
spherical Hopf algebra. The reason this abstraction is necessary is that
the category of representations of a Hopf algebra may be degenerate, and it
is necessary to take a non-degenerate quotient category to construct the
invariants.

This quotient is not the category of representations of
any finite dimensional Hopf algebra. The reason for this is that it is not
possible to assign a positive integer, the dimension, to each object which
is additive under direct sum and multiplicative under the tensor product.

First we recall the definition of a strict pivotal category given in
\Lcitemark Freyd\Nameand Yetter\Citebreak 1989\Rcitemark \Rspace{}. The
definition of a (relaxed) pivotal category  is
given in\Lspace \Lcitemark Barrett\Nameand Westbury\Citebreak 1993\Rcitemark
\Rspace{} and a similar definition is given in
\Lcitemark Freyd\Nameand Yetter\Citebreak 1992\Rcitemark \Rspace{}. A spherical
category is a pivotal category
which satisfies an additional condition.

In this paper we will only consider strict pivotal categories. There is
no loss of generality as it is shown in\Lspace \Lcitemark Barrett\Nameand
Westbury\Citebreak 1993\Rcitemark \Rspace{} that every
pivotal category is canonically equivalent to a strict pivotal category.
However the main examples of pivotal categories are categories of
representations of Hopf algebras and are not strict. The difference between
a pivotal category and a strict pivotal category is that some objects that
are
equal in a strict pivotal category are canonically isomorphic in a pivotal
category. In this section we denote any such canonical isomorphism by
$=$. These constructions can be extended to pivotal categories by putting
in the canonical isomorphism for each $=$.

\definition{Definition \propno} A category with strict duals consists of
a category $\Cal C$, a functor
$\otimes\colon{\Cal C}\times{\Cal C}\to {\Cal C}$, an object $e$ and a
functor $\spacehat\colon{\Cal C}\to {\Cal C}^{op}$. The conditions are that
$({\Cal C},\otimes ,e)$ is a strict monoidal category and
\roster
\item The functors $\spacehat\spacehat$ and $1$ are equal.
\item The objects $\hat e$ and $e$ are equal.
\item The functors ${\Cal C}\times{\Cal C}\to{\Cal C}$ which on objects are
given by $(a,b)\mapsto (a\otimes b)\sphat\ $ and
$(a,b)\mapsto \hat b\otimes \hat a$ are equal.
\endroster
\enddefinition

\definition{Definition \propno}
A strict pivotal category is a category with strict duals together with a
morphism $\epsilon (c)\colon e\to c\otimes \hat c$
for each object $c\in{\Cal C}$.

The conditions on the morphisms $\epsilon (c)$ are the following:
\roster
\item For all morphisms, $f\colon a\to b$, the following diagram commutes
$$\CD
e @>{\epsilon (a)}>> a\otimes \hat a\\
@V{\epsilon (b)}VV @VV{f\otimes 1}V\\
b\otimes \hat b @>>{1\otimes \hat f}> b\otimes \hat a
\endCD$$
\item For all objects $a$, the following composite is the identity map
of $\hat a$:
$$ \hat a = e\otimes \hat a @>{\epsilon (\hat a)\otimes 1}>>
(\hat a\otimes \ahathat)\otimes \hat a =
\hat a\otimes (a\otimes \hat a)\sphat @>{1\otimes {\hat \epsilon(a)}}>>
\hat a\otimes \hat e = \hat a $$
\item For all objects $a$ and $b$ the following composite is required to
be $\epsilon (a\otimes b)$:
$$e @>{\epsilon (a)}>> a\otimes \hat a =
a\otimes (e\otimes \hat a)
@>{1\otimes (\epsilon (b)\otimes 1)}>>
a\otimes ((b\otimes \hat b)\otimes \hat a) =
(a\otimes b)\otimes (a\otimes b)\sphat $$
\endroster
\enddefinition

The functor $\spacehat$ and the maps $\epsilon$ are not independent.  The maps
$\epsilon$ determine $\spacehat$.

\proclaim{Lemma \propno\name\dual}
In any pivotal category, for any morphism
$f\colon a\to b$ the following composite is $\hat f$:
$$\multline{\hat b} = {\hat b}\otimes e
@>{1\otimes\epsilon (a)}>> {\hat b}\otimes (a\otimes{\hat a})
@>{1\otimes (f\otimes 1)}>> {\hat b}\otimes (b\otimes{\hat a})\\
= \widehat{({\hat b}\otimes \bhathat)}\otimes {\hat a}
@>{\hat\epsilon (\hat b)\otimes 1}>> \hat e\otimes{\hat a}
= {\hat a}\endmultline$$
\endproclaim

\demo{Proof}  This follows directly from conditions (1) and (2) of
the preceding definition.
\enddemo

\definition{Definition \propno} Let $a$ be any object in a
pivotal category. Then the monoid $\End (a)$ has two trace maps,
$\tr_L,\tr_R\colon\End (a)\to\End (e)$.
In a pivotal category $\tr_L(f)$ is defined to be the composite
$$\multline
e @>{\epsilon (\hat a)}>>
\hat a\otimes \ahathat =
\hat a\otimes a @>{1\otimes f}>>
\hat a\otimes a =
(\hat a\otimes \ahathat)\sphat @>{\hat \epsilon(\hat a)}>> \hat e = e
\endmultline$$
and $\tr_R(f)$ is defined to be the composite
$$\multline
e @>{\epsilon (a)}>>
a\otimes \hat a @>{f\otimes 1}>>
a\otimes \hat a =
(a\otimes \hat a)\sphat @>{\hat \epsilon(a)}>>
\hat e = e
\endmultline$$
These are called trace maps because they satisfy $\tr_L(fg)=\tr_L(gf)$
and $\tr_R(fg)=\tr_R(gf)$.
\enddefinition

\definition{Definition \propno} A pivotal category is spherical if,
for all objects $a$ and all morphisms $f\colon a\to a$,
$$\tr_L(f)=\tr_R(f).$$
\enddefinition

An equivalent condition is that $\tr_L(f)=\tr_L(\hat f)$, for all
$f\colon a\to a$. For each object $a$ in a spherical category, its
quantum dimension is defined to be $\dim_q(a)=\tr_L(1_a)$. Thus,
$\dim_q(a)=\dim_q(\hat a)$.

Also, in a spherical category, $\tr_L(f\otimes g)=\tr_L(f).\tr_L(g)$
(where the product is in $\End(e)$)
for all $f\colon a\to a$ and all $g\colon b\to b$.

All spherical categories considered in the rest of this are additive.
This means that each $\Hom$ set is a finitely generated abelian group with
composition $\Z$-bilinear;
and that the data defining the spherical structure is compatible with the
additive structure. This means that $\otimes$ is $\Z$-bilinear and that
$\spacehat$ is $\Z$-linear.

In any additive monoidal category $\End (e)$ is a commutative ring
(see\Lspace \Lcitemark Kelly\Nameand Laplaza\Citebreak 1980\Rcitemark
\Rspace{})  and we denote this ring by $\ring$.
In particular each of the above trace maps takes values in this ring.
It follows that an additive monoidal category is $\ring$-linear.
We need to assume some further conditions. These are that
$\ring$ is a field and that each set of morphisms is a finite dimensional
vector space over $\ring$. In this paper an additive spherical category
means a spherical category in which these conditions are satisfied.

The main examples of additive spherical categories arise
as the category of representations of a Hopf algebra with some additional
structure. This is discussed in section 6.

\example{Example \propno} An example of a spherical category
which cannot be regarded as a category whose objects are finite
dimensional vector spaces is given by taking the free $\Z [\delta ,z]$-linear
category on the category of oriented framed tangles and then taking the
quotient by the well-known skein relation for the HOMFLY polynomial.
This is a spherical category and for each pair of objects $X$ and
$Y$, $\Hom (X,Y)$ is a finitely generated free module. This example
saisfies all the conditions for an additive spherical category
except that $\ring$ is not a field. However the objects
cannot be taken to be finitely generated modules unless $z$ is a quantum
integer.
\endexample
\nextfigure\trace
\goodbreak\midinsert \centerline {

\epsfbox{}
}
\figno
\endinsert

Strict pivotal categories are discussed  in\Lspace \Lcitemark Freyd\Nameand
Yetter\Citebreak 1989\Rcitemark \Rspace{}, where it is
shown
that the category of oriented planar graphs up to isotopy and with labelled
edges is a strict pivotal category. Similar constructions are discussed
\Lcitemark Joyal\Nameand Street\Citebreak 1991\Rcitemark \Rspace{},\Lspace
\Lcitemark Freyd\Nameand Yetter\Citebreak 1992\Rcitemark \Rspace{},\Lspace
\Lcitemark Reshetikhin\Nameand Turaev\Citebreak 1990\Rcitemark \Rspace{}.
The following is an informal
statement of the result needed in the discussion in this paper. This result
is not required for the proofs in this paper, as an algebraic proof can
always be given in place of a diagrammatic proof. However it is important
for an understanding of the proofs.

Given a trivalent planar graph with the following data attached:
\roster
\item An orientation of each edge
\item A distinguished edge at each trivalent vertex
\item A map from edges to objects
\item A map from vertices to morphisms
\endroster
then this graph can be evaluated to give a morphism. The
relations for a pivotal category imply that this evaluation depends only on
the isotopy class of the graph, where the attached data is carried along
with the isotopy.

The sphere $S^2$ can be regarded as the plane with the point at infinity
attached, and so a planar graph can also be thought of as a graph embedded
in $S^2$. Spherical categories are pivotal categories which satisfy an
extra condition, the equality of left and right trace. This condition
implies that the spherical category determines an invariant of isotopy
classes of closed graphs on the sphere.
There is an isotopy of the sphere which takes a closed graph
of the form of Figure \trace\ to the graph obtained by closing $M$ in a loop
to the left. This isotopy moves the loop in Figure \trace\  past the point at
infinity. Taken together with planar isotopies, such an operation on planar
graphs generates all the isotopies on the sphere.

\definition{Definition \propno} For any two objects $a$ and $b$ there
is a bilinear pairing
$$\Theta\colon\Hom (a,b)\times\Hom (b,a)\to\ring$$
defined by $\Theta (f,g)=\tr_L(fg)=\tr_L(gf)$.
\enddefinition

\definition{Definition \propno}
An additive spherical category is non-degenerate if, for all objects $a$ and
$b$,
the pairing $\Theta$ is non-degenerate. \enddefinition

The next theorem shows that every additive spherical category has a
natural quotient which is a non-degenerate spherical category.
\proclaim{Theorem \propno \name\quotient} Let ${\Cal C}$ be an additive
spherical category. Define the additive subcategory ${\Cal J}$ to have the same
set of
objects and morphisms defined by
$$\Hom_{\Cal J}(c_1,c_2)=\{f\in \Hom_{\Cal C}(c_1,c_2): \tr_L (fg)=0\qquad
\text{for all $g\in \Hom_{\Cal C}(c_2,c_1)$}\}$$
Then ${\Cal C}/{\Cal J}$ is a non-degenerate additive spherical category.
\endproclaim
\demo{Proof} It is clear that ${\Cal J}$ is closed under composition on
either side by arbitrary morphisms in ${\Cal C}$. Hence the quotient is an
additive category. It is also clear that $\hat f\in {\Cal J}$ if and only if
$f\in {\Cal J}$ and so the functor $\spacehat$ is well-defined on the
quotient. The functor $\otimes$ is well-defined on the quotient since
$f\in {\Cal J}$ implies $f\otimes g_1\in {\Cal J}$ and
$g_2\otimes f\in {\Cal J}$ for arbitrary morphisms in ${\Cal C}$.
This follows from the observation that $\tr_L(f\otimes g)=
\tr_L(f).\tr_L(g)$ which uses the spherical condition.

The morphisms  $\epsilon (a)$ are taken to be
the images in the quotient of the given morphisms in ${\Cal C}$. The
conditions on this structure which imply that this quotient is spherical
follow from the same conditions in ${\Cal C}$.

Each pairing $\Theta$ is non-degenerate by construction.
\enddemo

An extra condition on the spherical category is required for the
piecewise-linear invariance of the partition function of a natural
simplicial field theory. Similar conditions have been considered by
\Lcitemark Reshetikhin\Nameand Turaev\Citebreak 1991\Rcitemark
\Rspace{},\Lspace \Lcitemark Walker\Citebreak 1990\Rcitemark \Rspace{},\Lspace
\Lcitemark Turaev\Nameand Wenzl\Citebreak 1993\Rcitemark \Rspace{},\Lspace
\Lcitemark Turaev\Citebreak 1992\Rcitemark \Rspace{}
and\Lspace \Lcitemark Yetter\Citebreak 1993\Rcitemark \Rspace{}.
An object $a$ is called non-zero if the ring $\End(a)\ne 0$.

\definition{Definition \propno\name\semisimple} A semisimple spherical
category is an additive, non-degenerate, spherical category such that
there exists a set of inequivalent non-zero objects, $J$, such that for any
two objects $x$ and $y$, the natural map given by composition,
$$\oplus_{a\in J}\Hom (x,a)\otimes\Hom (a,y)\to\Hom(x,y),$$
is an isomorphism.
\enddefinition

An object $a$ is called simple if $\End(a)\cong \ring$.

The following lemma shows that the set $J$ is essentially fixed by
the category.

\proclaim{Lemma \propno} Every simple object is isomorphic to a unique
element of $J$, and every element of $J$ is simple.
\endproclaim

\demo{Proof}
In the formula
$$\oplus_{a\in J}\Hom (x,a)\otimes\Hom (a,x)\cong\End(x)$$
first consider $x$ to be an element of $J$. Then by counting dimensions,
one has that $\End(x)\cong \ring$.

Now consider the same formula with $x$ any simple object.
Again by counting dimensions, only one of the terms on the left is non-zero.
For this $a\in J$, $\Hom(x,a)\cong\Hom(a,x)\cong\ring$. Thus there
are elements $f\in\Hom(x,a)$, $g\in\Hom(a,x)$ such that $fg=id_x$.
{}From this it follows that $gf\in \End(a)$ is an idempotent and is not zero.
But $\End(a)\cong \ring$, and so $gf=id_a$. This shows that $x$ is
isomorphic to $a\in J$.
\enddemo

\comment For any objects $x,y$, the non-degeneracy condition shows
that $\Hom(x,y)$ and $\Hom(y,x)$ are dual and so have the same dimension.
\endcomment

\definition{Definition \propno} A semisimple spherical category is called
finite if the set of isomorphism classes of simple objects is finite.
\enddefinition

\definition{Definition \propno} The dimension $K$ of a finite semisimple
spherical category is defined by the formula
$$K=\sum_{a\in J}\dim_q^2(a)$$
for some choice $J$ of one object in each isomorphism class of simple
objects. The dimension
is independent of this choice.\enddefinition

\proclaim{Lemma \propno\name\dimensions}
For each pair of objects $(a,b)$ in a semisimple spherical category,
$$\dim_q (a) \dim_q (b)=\sum_{c\in J}\dim_q (c)\dim\Hom(c,a\otimes b).$$
\endproclaim

\demo{Proof} The left hand side is equal to $\tr 1_{a\otimes b}$. The lemma
follows from the application of the semisimple condition of definition
\semisimple{} with $x=y=a\otimes b$, and some linear algebra.\enddemo

\head\headno Symmetries of simplicial invariants \endhead

In this section, we define the data for a state sum model given a strict
non-degenerate
spherical category $\Cal C$. Then we show that the simplicial invariant of
labelled manifolds has the property that it depends only on the isomorphism
class of the labelling of each edge. Then it is shown that the invariant
depends only on the underlying combinatorial structure of the simplicial
complex.

The data for a state sum model is constructed as follows. The label set
$I$ is the set of simple objects in the category.
For each ordered triple $(a,b,c)$ of labels, the vector space $H(a,b,c)$ is
defined to be $\Hom(b,a\otimes c)$\nextfigure\Y\ (See Figure~\Y).

\goodbreak\midinsert
\centerline{\epsfbox{y.eps}}
\figno\endinsert

\nextfigure\simplex

For the partition function of the tetrahedron A labelled by $e_{01}=a$,
$e_{02}=b$, $e_{12}=c$, $e_{23}=d$, $e_{13}=e$, $e_{03}=f$, first define
a linear functional on the space
$$ \Hom (d\otimes c,e)\otimes\Hom (f,d\otimes b)\otimes
\Hom (e\otimes a,f)\otimes\Hom (b,c\otimes a).$$
The linear functional is defined to be
$$\alpha\otimes\beta\otimes\gamma\otimes\delta\mapsto
\tr_L(\beta (1\otimes\delta )(\alpha\otimes 1)\gamma )$$
(Figure~\simplex). This linear functional determines a unique linear map
$$\left\{\matrix a&b&c\\d&e&f\endmatrix\right\}_+\colon
\Hom(f,d\otimes b)\otimes\Hom(b,c\otimes a)\to
\Hom(e,d\otimes c)\otimes\Hom(f,e\otimes a)$$
using the non-degenerate pairings $\Hom^*(d\otimes c,e)\cong
\Hom(e,d\otimes c)$, and $\Hom^*(e\otimes a,f)\cong
\Hom(f,e\otimes a)$.

For the partition function of $-A$ labelled in the same way, the linear
functional on
$$ \Hom (e,d\otimes c)\otimes\Hom (d\otimes b,f)\otimes
\Hom (f,e\otimes a)\otimes\Hom (c\otimes a,b)$$
defined by
$$\alpha\otimes\beta\otimes\gamma\otimes\delta\mapsto
\tr_L(\gamma (\alpha\otimes 1) (1\otimes\delta )\beta).$$
likewise determines a unique linear map
$$\left\{\matrix a&b&c\\d&e&f\endmatrix\right\}_-\colon
\Hom(e,d\otimes c)\otimes\Hom(f,e\otimes a)
\to
\Hom(f,d\otimes b)\otimes\Hom(b,c\otimes a)$$

\goodbreak\midinsert
\centerline{

\epsfbox{}}
\figno\endinsert

\definition {Definition \propno\name\naturaliso} Given isomorphisms
$\phi_a\colon a\to a'$,
$\phi_b\colon b\to b'$ and $\phi_c\colon c\to c'$, then there is an
induced isomorphism
$$\Hom(b,a\otimes c) \to \Hom(b',a'\otimes c')$$
given by $\alpha\mapsto \phi_b^{-1}\alpha(\phi_a\otimes \phi_c)$.
\enddefinition

\proclaim{Lemma \propno} Given any ordered 6-tuple of elements of $I$,
$(a,b,c,d,e,f)$, and an ordered 6-tuple of isomorphisms,
$(\phi_a,\phi_b,\phi_c,\phi_d,\phi_e,\phi_f)$, where
$\phi_a\colon a\to a^\prime$,$\ldots $,$\phi_f\colon f\to f^\prime$,
then the following diagram commutes:
$$\CD \Hom(f,d\otimes b)\otimes\Hom(b,c\otimes a)
@>>>
\Hom(f',d'\otimes b')\otimes\Hom(b',c'\otimes a')\\
@V{\left\{\matrix a&b&c\\d&e&f\endmatrix\right\}_+}VV
@VV{\left\{\matrix a^\prime&b^\prime&c^\prime\\d^\prime&e^\prime&f^\prime
\endmatrix\right\}_+}V\\
\Hom(e,d\otimes c)\otimes\Hom(f,e\otimes a)
@>>>
\Hom(e',d'\otimes c')\otimes\Hom(f',e'\otimes a')\\
\endCD$$
Also, the diagram for the opposite orientation commutes:
$$\CD \Hom(f,d\otimes b)\otimes\Hom(b,c\otimes a)
@>>>
\Hom(f',d'\otimes b')\otimes\Hom(b',c'\otimes a')\\
@A{\left\{\matrix a&b&c\\d&e&f\endmatrix\right\}_-}AA
@AA{\left\{\matrix a^\prime&b^\prime&c^\prime\\d^\prime&e^\prime&f^\prime
\endmatrix\right\}_-}A\\
\Hom(e,d\otimes c)\otimes\Hom(f,e\otimes a)
@>>>
\Hom(e',d'\otimes c')\otimes\Hom(f',e'\otimes a')\\
\endCD
$$\endproclaim

\demo{Proof} First, in the diagram
$$\CD
\Hom(e,d\otimes c)
@>\alpha\mapsto\tr_L(\alpha\placemarker)>>
\Hom^* (d\otimes c,e)
\\
@VVV @VVV\\
\Hom(e',d'\otimes c')
@>\alpha\mapsto\tr_L(\alpha\placemarker)>>
\Hom^*(d'\otimes c' ,e')
\endCD$$
the horizontal arrows are defined by the pairings,
the left vertical arrow by the induced isomorphism of the previous definition,
and the right vertical arrow by the adjoint of the map
$\beta\mapsto(\phi_e\otimes\phi_a)\beta\phi_f^{-1}$. This diagram, and
a similar diagram obtained by replacing $e,d,c$ with $f,e,a$, commute.
These diagrams are used to compute the action of the isomorphisms
of the statement of the lemma on the linear functionals in the definition
of the partition function.

The first diagram in the statement of the lemma commutes as a consequence of
the identity
$$\multline
\tr_L \phi_f^{-1}\beta(\phi_d\otimes\phi_b)
(1\otimes(\phi_b^{-1}\delta(\phi_c\otimes\phi_a)))
(((\phi_d\otimes\phi_c)^{-1}\alpha\phi_e)\otimes 1)
(\phi_e\otimes\phi_a)^{-1}\gamma\phi_f\\
=\tr_L(\beta (1\otimes\delta )(\alpha\otimes 1)\gamma ).\endmultline$$
The proof that the second diagram commutes is similar.\enddemo

\proclaim{Proposition \propno\name\naturality} Let $M$ be a closed simplicial
manifold. Let $l_1$ and $l_2$ be two labellings such that the two labels
associated to any edge are isomorphic. Then, $Z(M,l_1)=Z(M,l_2)$.
\endproclaim

\demo{Proof} According to the previous lemma, the map $V(M)\to V(M)$ is
conjugated by the induced isomorphism on the state space of each triangle.
The invariant $Z(M)$ is the trace of this map and is invariant under
conjugation by a linear map.
\enddemo

Next, we determine the behaviour of the simplicial invariant $Z(M)$ under
combinatorial maps. For this, it is necessary to use the properties of
duals in the spherical category.

\definition{Definition \propno} Let $f\colon M\to N$ be a combinatorial
isomorphism of simplicial complexes. Let $e$ be any edge of $M$, labelled
by $a$, and let $b$ be the label of edge $f(e)$ in $N$. Then
$f$ is compatible with these labellings if $b=a$ in the case that $f$ preserves
the orientation of the edge, and $b=\hat a$ in the case that $f$ reverses
the orientation.\enddefinition

Note that, given $f$ and a labelling of $M$, there is a unique compatible
labelling of $N$.

{}Now the properties of the state space of a triangle under combinatorial
isomorphisms are described. The combinatorial isomorphisms are just
permutations in $S_3$. For the standard triangle $T(a,\hat b,c)$, labelled by
$\partial_0T\mapsto a$, $\partial_1T\mapsto \hat b$, $\partial_2T\mapsto c$,
the labelling is permuted by $(a,b,c)\mapsto \sigma^+(a,b,c)$ for an even
permutation $\sigma^+$, and $(a,b,c)\mapsto \sigma^-(\hat a,\hat b,\hat c)$
for an odd permutation $\sigma^-$. For this reason, it is more convenient to
use the
notation $V(a,b,c)$=$H(a,\hat b,c)$ for the state space of a labelled
triangle when the symmetry properties are considered.

There is a canonical map $V(a,b,c)\to V(b,c,a)$, i.e.,
$$ \Hom (\hat b ,a\otimes c)\to
\Hom (\hat c ,b\otimes a),$$
defined by mapping $f\colon\hat b\to a\otimes c$ to the following composite:
$$\multline
\hat c @>=>>
e\otimes\hat c @>{\epsilon (b)\otimes 1}>>
b\otimes\hat b\otimes \hat c   @>{1\otimes f\otimes 1}>>
b\otimes a\otimes c\otimes\hat c
 @>{1\otimes 1\otimes{\hat\epsilon(c)}}>>
b\otimes a\otimes e @>=>>
b\otimes a.
\endmultline$$
\nextfigure\thirdturn
This corresponds to the graph in Figure~\thirdturn.
\goodbreak\midinsert
\centerline{\epsfbox{}}
\figno\endinsert

There is also a canonical pairing $V(a,b,c)\times
V(\hat c,\hat b,\hat a)\to\ring$, i.e.,
$$\Hom (\hat b ,a\otimes c)\times\Hom (b,\hat c\otimes\hat a)\to\ring.$$
Let $f\colon\hat b\to a\otimes c$ and
$g\colon b\to\hat c\otimes\hat a$ then the pairing is defined by
$$\left< f,g\right>=\tr_L(f\hat g)$$

\nextfigure\pairing
Equivalently, it is determined by the closed tangle in
 Figure~\pairing.
\goodbreak\midinsert
\centerline{\epsfbox{}}
\figno\endinsert

\definition{Definition \propno}
For every ordered triple, $(a,b,c)$, of elements of $I$
and every even permutation $\sigma^+\in S_3$, there is an isomorphism
$$\theta(\sigma^+)\colon V(a,b,c)\to V{\sigma^+(a,b,c)}.$$
For $(a,b,c)\mapsto (b,c,a)$, this is the canonical map just defined.
Repeating this gives the isomorphism for $(a,b,c)\mapsto (c,a,b)$. The
identity is associated to the identity.

For every ordered triple, $(a,b,c)$, of elements of $I$,
and every odd permutation $\sigma^-\in S_3$, there is a non-degenerate
pairing, $\left< -,-\right>_{\sigma^-}$,
$$V(a,b,c)\otimes V{\sigma^-(\hat a,\hat b,\hat c)}\to\ring .$$
For $\sigma^-$ the odd permutation $(a,b,c)\mapsto (c,b,a)$, this is the
pairing defined above. The pairings for the other two odd permutations can
be defined by the formula
$\left< v_1,v_2\right>_{\sigma^-}
=\left< v_1,\theta(\sigma^+) v_2\right>_{\sigma^-\sigma^+}$.
\enddefinition

\proclaim{Lemma \propno\name\conditions}
For all even permutations $\sigma^+_1,\sigma^+_2$, odd permutations
$\sigma^-$,
labels $a$, $b$ and $c$ and all $v_1\in V(a,b,c)$ and $v_2\in V{\sigma^-
(\hat a,\hat b,\hat c)}$ :
$$\align
\theta (\sigma_1^+\sigma_2^+)&=\theta (\sigma_1^+)\theta (\sigma_2^+) \\
\left< v_1,v_2\right>_{\sigma^-}
&=\left< v_1,\theta(\sigma^+) v_2\right>_{\sigma^-\sigma^+} \\
\left< v_1,v_2\right>_{\sigma^-} &=\left< v_2,v_1\right>_{\sigma^-}
\endalign$$
Also, the pairings are non-degenerate bilinear forms.\endproclaim

\demo{Proof}
The first two relations follow from the fact that the
following composite is the identity map.
$$V(a,b,c)@>>>V(b,c,a)@>>>V(c,a,b)@>>>V(a,b,c)$$
\nextfigure\vortex
This condition is the relation shown in Figure~\vortex\  which is
satisfied in any pivotal category, by Lemma \dual .

\goodbreak\midinsert
\centerline{
\epsfbox{
}}
\figno\endinsert

The pairings are non-degenerate since the spherical category is
non-degenerate and $\spacehat$ is an isomorphism on spaces of morphisms.
The pairing $\tr_L(f\hat g)$ is symmetric since
$$\tr_L(g\hat f)=\tr_L(\widehat{g\hat f})=\tr_L(f\hat g).$$
The symmetry of the other pairings is equivalent to the relations
$$\tr_L((\theta(\sigma^+)v_2)\hat v_1)=\tr_L((\theta(\sigma^+)v_1)\hat v_2).$$
This follows from the relations in a spherical category, as can be seen
by the isotopy equivalence of the corresponding diagrams.
\enddemo

The union of the spaces $\{V(a,b,c)\coprod V^*(a,b,c)\mid (a,b,c)\in
I\times I\times I\}$
forms a vector bundle over $I\times I\times I\times\{\pm\}$.
The permutation group, $S_3$, acts on the base space by
$$\align
\sigma^+\colon(a,b,c,\pm)&\mapsto (\sigma^+(a,b,c),\pm)\\
\sigma^-\colon(a,b,c,\pm)&\mapsto (\sigma^-(\hat a,\hat b,\hat c),\mp).
\endalign$$

\definition {Definition \propno\name\action} For each triple of labels,
$(a,b,c)$, and each
permutation $\sigma^+$ or $\sigma^-$ there are linear isomorphisms
$$\align V(a,b,c)&@>\theta(\sigma^+)>>V\sigma^+(a,b,c)\\
V^*(a,b,c)&@>\theta^{*-1}(\sigma^+)>>V^*\sigma^+(a,b,c)\endalign$$
if $\sigma^+$ is even, and
$$\align V(a,b,c)&@>\theta(\sigma^-)>>V^*\sigma^-(\hat a,\hat b,\hat c)\\
V^*(a,b,c)&@>\theta^{*-1}(\sigma^-)>>V\sigma^-(\hat a,\hat b,\hat c)
\endalign$$
if $\sigma^-$ is odd. The maps $\theta(\sigma^-)$ are defined using the
pairings. \enddefinition

\proclaim{Lemma \propno}  These linear maps determine an
action of the group $S_3$ on this vector bundle, or, in other words,
this is an $S_3$-equivariant vector bundle. Furthermore,the action of any
permutation on elements of $V^*(a,b,c)$
is the adjoint of the inverse of the action on $V(a,b,c)$.
\endproclaim

\demo{Proof} These are equivalent to the conditions in lemma \conditions.
\enddemo

\proclaim{Theorem \propno\name\symmetry} Let $f\colon M\to N$ be a
combinatorial isomorphism of labelled manifolds.
Then the simplicial invariants are equal, $Z(M)=Z(N)$.
\endproclaim

\demo{Proof} Let $V(M)$ and $V(N)$ be the vector spaces
described in definition \simplicialinvariant. For each triangle in $M$,
consider the restriction of $f$ to this triangle. There is a element
$\sigma\in S_3$ defined by the unique decomposition of this map into a
permutation followed by the simplicial map of the triangle to its image in
$N$.

There is a map
$$V(M)\otimes V^*(M)\to V(N)\otimes V^*(N)$$
which is defined by taking the tensor product over the set of triangles
of the maps
$$\theta(\sigma)\otimes\theta^{*-1}(\sigma)$$
for each triangle, followed by an iteration of the standard twist $P$
which rearranges the factors in the range of this map to coincide
with $V(N)\otimes V^*(N)$, as in definition \simplicialinvariant.
Since the action of $\sigma$ on $V^*(e_1,e_2,e_3)$ is the inverse
of the adjoint of the action on $V(e_1,e_2,e_3)$, it follows that the diagram
$$\CD  V(M)\otimes V^*(M)@>>>V(N)\otimes V^*(N)\\
@VVV @VVV\\
\ring @= \ring \endCD$$ in which the vertical maps are the canonical pairings,
commutes.

To complete the proof of the theorem, it remains to show that the map
$V(M)\to V(M)$
whose trace is $Z(M)$ is preserved under this mapping. That is, that this
element of $V(M)\otimes V^*(M)$ is mapped to the corresponding element of
$V(N)\otimes V^*(N)$. According to definition \simplicialinvariant,
each of these elements is the tensor product of partition functions for each
tetrahedron. Thus it is sufficient to show that the partition function of
the standard tetrahedron is preserved under a combinatorial mapping.
This is demonstrated by the next lemma.
\enddemo

\definition{Definition \propno} Let $T$ be an oriented labelled simplicial
surface. Then the state space of $T$ is defined to be the tensor product
over the set of triangles of the state space for each oriented triangle.
\enddefinition

Let $T\to U$ be an orientation-preserving combinatorial isomorphism of
oriented labelled simplicial surfaces which is compatible with the
labellings. Then there is a linear isomorphism from the state space
of $T$ to the state space of $U$. On each triangle in $T$ an element of $S_3$
is determined such that the combinatorial map is a permutation followed by a
simplicial map. The linear isomorphism is defined by taking the tensor
product over triangles of the linear isomorphisms of definition \action.
This tensor product is composed with the unique iterate of the twist map
$P$ which has its range the state space of $U$.

\proclaim{Lemma \propno} The partition function of the standard labelled
tetrahedra $A$ and $-A$ are elements of the state spaces of their boundary.
Let $\Sigma\in S_4$ be an even permutation. The element $\Sigma$ determines
combinatorial maps $A\to A'$ and $-A\to -A'$, where $A'$ is labelled by
a compatible labelling $\{e'_{ij}\}$. Under the linear map of state spaces,
$$\left\{\matrix e_{01}&e_{02}&e_{12}\\e_{23}&e_{13}&e_{03}
\endmatrix\right\}_+
\mapsto
\left\{\matrix e'_{01}&e'_{02}&e'_{12}\\e'_{23}&e'_{13}&e'_{03}
\endmatrix\right\}_+$$
and
$$\left\{\matrix e_{01}&e_{02}&e_{12}\\e_{23}&e_{13}&e_{03}
\endmatrix\right\}_-
\mapsto
\left\{\matrix e'_{01}&e'_{02}&e'_{12}\\e'_{23}&e'_{13}&e'_{03}
\endmatrix\right\}_-.$$
If $\Sigma$ is an odd permutation, then
$$\left\{\matrix e_{01}&e_{02}&e_{12}\\e_{23}&e_{13}&e_{03}
\endmatrix\right\}_+
\mapsto
\left\{\matrix e'_{01}&e'_{02}&e'_{12}\\e'_{23}&e'_{13}&e'_{03}
\endmatrix\right\}_-$$
and
$$\left\{\matrix e_{01}&e_{02}&e_{12}\\e_{23}&e_{13}&e_{03}
\endmatrix\right\}_-
\mapsto
\left\{\matrix e'_{01}&e'_{02}&e'_{12}\\e'_{23}&e'_{13}&e'_{03}
\endmatrix\right\}_+.$$
\endproclaim

\demo{Proof} The first statement follows from the isomorphism $\Hom(X,Y)
\cong X^*\otimes Y$ for vector spaces $X,Y$.

In the definition of the partition function of the tetrahedron, each
factor \break $\Hom(\hat b,a\otimes c)$ in the tensor product is identified
with
$\Hom^*(a\otimes c,\hat b)$, using the non-degenerate symmetric pairing
$(\alpha,\beta)
\mapsto \tr_L(\alpha\beta)$ in the spherical category. Using these
isomorphisms, the action of the odd and even permutations can be computed
by the following commutative diagrams:
$$\CD
\Hom(\hat b,a\otimes c)
@>\theta(\sigma^+)>>
\Hom (\hat c,b\otimes a)
\\
@V\alpha\mapsto\tr_L(\alpha\placemarker)VV
@VV\alpha\mapsto\tr_L(\alpha\placemarker)V\\
\Hom^*(a\otimes c,\hat b)
@>\phi^*>>
\Hom^*(b\otimes a ,\hat c )\endCD$$
in which $\phi\colon\beta\mapsto \widehat{\theta(\sigma^+)\hat\beta}$, and
$$\CD
\Hom(\hat b,a\otimes c)
@>\theta(\sigma^-)>>
\Hom^* (b,\hat c\otimes \hat a)
\\
@V\alpha\mapsto\tr_L(\alpha\placemarker)VV @|\\
\Hom^*(a\otimes c,\hat b)
@>>(\spacehat)^*>
\Hom^* (b,\hat c\otimes \hat a).\endCD$$
The map $(\spacehat)^*$ is the adjoint of the linear map $\spacehat$.

{}From these diagrams, the action of the elements of $S^4$ on the linear
functionals in the definition of the partition function can be computed.
The maps  $\phi$ and $\theta(\sigma^+)$ correspond, as
diagrams, to rotations by one third of a turn, and $\spacehat$ corresponds
to one half of a turn.
For even permutations in $S_4$, the symmetry property of the partition
function follows from the fact that
any even permutation of the vertices of a tetrahedron can be extended to an
isotopy of the sphere. The definition of spherical category was constructed
to give invariants of isotopy classes of graphs on the sphere.
For odd elements of $S_4$, the symmetry property follows from the fact
that any odd permutation of the vertices of a tetrahedron can be extended
to an isotopy of the sphere which takes the tetrahedron to its image under
some fixed reflection in a diameter. The diagrams corresponding to $A$
and $-A$ differ by such a reflection.

Alternatively, the symmetry property can be checked algebraically.
As an example, consider the odd permutation $(0,1,2,3)\mapsto(3,1,2,0)$.
The state space of $A$ is
$$H(e_{23},e_{13},e_{12} )\otimes
H^*(e_{23},e_{03},e_{02})\otimes H(e_{13},e_{03},e_{01})
\otimes H^*(e_{12},e_{02},e_{01})$$
and the state space of $A'$
$$H^*(e'_{23},e'_{13},e'_{12} )\otimes
H(e'_{23},e'_{03},e'_{02})\otimes H^*(e'_{13},e'_{03},e'_{01})
\otimes H(e'_{12},e'_{02},e'_{01})$$
The linear map of state spaces is
$$x\otimes y\otimes z\otimes t\mapsto
\theta^{*-1}(\sigma^{-1})t\otimes\
\theta^{*-1}(\tau)y\otimes
\theta(\tau)z\otimes
\theta(\sigma)x,$$
where $\sigma$ is the permutation $(0,1,2)\mapsto (1,2,0)$,
and $\tau$ is the permutation $(0,1,2)\mapsto (2,1,0)$.
As a map of linear functionals, this is the adjoint of the map
$$\alpha\otimes\beta\otimes\gamma\otimes\delta\mapsto
\theta(\sigma)\delta\otimes\
\hat \beta\otimes
\hat\gamma\otimes
\widehat{\theta(\sigma)\hat\alpha}.$$
The symmetry property is equivalent to the identity
$$\tr_L\left(\hat\gamma\bigl(\theta(\sigma)\delta\otimes 1\bigr)
\bigl(1\otimes \widehat{
\theta(\sigma)\hat\alpha}\bigr)
\hat\beta\right)
=\tr_L\bigl(\beta \bigl(1\otimes\delta \bigr)\bigl(\alpha\otimes 1\bigr)
\gamma \bigr)$$
which holds in any spherical category.
\enddemo

\head\headno Piecewise-linear manifolds \endhead

The aim of this section is to give a finite set of moves on the
triangulations of a 3-manifold such that any two triangulations are
related by a finite sequence of these moves.

\definition{Definition \propno} If $\sigma$ is a simplex in a complex $K$
then the star of $\sigma$ is the union of simplices in $K$
which contain $\sigma$.
\enddefinition
\definition{Definition \propno} If $\sigma$ is a simplex in a complex $K$
then the link of $\sigma$ is the union of all the simplices
in the star of $\sigma$ which do not meet $\sigma$.
\enddefinition
If $\sigma$ is a $k$-simplex in an $n$-manifold then the link of $\sigma$
is a sphere of dimension $n-k-1$. Also, the star of $\sigma$ is the
join of $\sigma$ and the link of $\sigma$.
\definition{Definition \propno} Let $\sigma$ be a $k$-simplex in a
$n$-manifold.
The cone on the star of $\sigma$ is an $(n+1)$-ball. The boundary
of this $(n+1)$-ball consists of two $n$-balls, $B_1$ and $B_2$,
one of which, say $B_1$, is the star of $\sigma$.
Stellar subdivision on $\sigma$ consists of removing $B_1$ from
the manifold and replacing it with $B_2$.
\enddefinition
\definition{Definition \propno} A stellar move is a stellar subdivision on a
simplex or the inverse of a stellar subdivision on a simplex.
\enddefinition
\nextfigure\stellars
The stellar move on a 2-simplex in a 3-manifold is drawn in Figure%
{}~\stellars.
\goodbreak\midinsert
\centerline{\epsfbox{stellar2.eps}}
\figno\endinsert
\nextfigure\stellart
The stellar move on a 3-simplex in a 3-manifold is drawn in Figure%
{}~\stellart  .
\goodbreak\midinsert
\centerline{\epsfbox{stellar3.eps}}
\figno\endinsert
The following theorem is given for manifolds in\Lspace \Lcitemark
Alexander\Citebreak 1930\Rcitemark \Rspace{}, and
generally in\Lspace \Lcitemark Glaser\Citebreak 1970\LIcitemark{}, Chapter II
\S D, Theorem II.17\RIcitemark \Rcitemark \Rspace{}.
\proclaim{Theorem \propno} Two finite simplicial complexes are
piecewise-linear homeomorphic if and only if they are related
by a finite sequence of stellar moves.
\endproclaim

In general there are infinitely many different stellar subdivisions.
More precisely there is a stellar subdivision of order $k$ in an
$n$-manifold for each triangulation of the sphere $S^{n-k-1}$.
This problem first arises in dimension 3 where there are infinitely
many stellar subdivisions of order 1. We will now introduce for
each $n$ a finite list of moves and show that in dimension 3 these
are enough.

Let $\sigma^n$ be an $n$-simplex. For any $p$ and $q$, the complexes
$\partial\sigma^p*\sigma^q$ and $\sigma^p*\partial\sigma^q$
are triangulations of the solid ball, $B^{p+q}$, and have the same
boundary, namely $\partial\sigma^p*\partial\sigma^q$. Also the union of
$\partial\sigma^p*\sigma^q$ and $\sigma^p*\partial\sigma^q$
along their boundary is $\partial\sigma^{p+q+1}$ and all decompositions
of this sphere into two disks arise this way.

\definition{Definition \propno} For any $k$ such that
$0\leqslant k\leqslant n$,
if $X$ is any $n$-manifold with an identification
of a boundary component with $\partial\sigma^k*\partial\sigma^{n-k}$
then $X\cup\partial\sigma^k*\sigma^{n-k}$ is said to be obtained from
$X\cup\sigma^k*\partial\sigma^{n-k}$ by an elementary move of order $k$.
\enddefinition

\nextfigure\be

\example{Example \propno}
 Figure~\be\ shows an elementary move of order 2 in a 3-manifold.
\goodbreak\midinsert
\centerline{\epsfbox{be.eps}}
\figno\endinsert
On the left-hand side there are two tetrahedra with a common horizontal
face. On the right there are three tetrahedra with a common vertical edge.
\endexample

Note that elementary moves have the following properties:
\roster
\item An elementary move of order $n$ is the same as a
stellar subdivision of order $n$.
\item The inverse of an elementary move of order $k$ is an
elementary move of order $n-k$.
\item An elementary move on a manifold with boundary does not
change the triangulation of the boundary.
\endroster

\proclaim{Lemma \propno} A stellar subdivision of order $(n-1)$ in a  closed
$n$-manifold can be written as a composite of two elementary moves of
orders $n$ and $n-1$. \endproclaim
\demo{Proof} Every $(n-1)$-simplex, $\sigma$, in a closed $n$-manifold
is a face of exactly two $n$-simplices. The link of $\sigma$ consists
of two points, say $N$ and $S$.

Then do an elementary move of order $n$ on $N\sigma$
introducing a new vertex $O$. This is, of course, the same as
doing stellar subdivision on $N\sigma$. Now do an
elementary move of order $n-1$ on the two $n$-simplices
$O\sigma$ and $S\sigma$. The effect of this
is to introduce the edge $OS$ and gives the same complex as
stellar subdivision on $\sigma$. \enddemo

\proclaim{Lemma \propno} A stellar subdivision of order $(n-2)$ in a  closed
$n$-manifold can be written as a finite sequence of elementary moves.
\endproclaim

\demo{Proof} It is sufficient to prove this for $n=3$. The result for $n>3$
is proved in exactly the same manner, by joining every complex mentioned to
a fixed $(n-4)$-simplex.

Let $NS$ be the vertices
of an edge which is in $p$ tetrahedra. Label the vertices of these
$p$ tetrahedra so that the vertices of these tetrahedra are
$$NSE_iE_{i+1}\qquad\text{for $1\leqslant i\leqslant p$}.$$
This gives a triangulation of the 3-ball which looks like
an orange with $p$-segments. Doing stellar subdivision on the edge
$NS$ gives a triangulation of the 3-ball with $2p$ tetrahedra.
This is like slicing the orange in half, cutting each segment in half.

In order to obtain this complex from the original one by elementary moves
first do an elementary move of order $4$ on the tetrahedron
$NSE_1E_2$. This introduces a new vertex which we label $O$. This
gives the right number of vertices but only $p+3$ tetrahedra.
Now for $j=2,3,\ldots n$ do an elementary move of order 3 on the
two tetrahedra $ONSE_j$ and $NSE_jE_{j+1}$. This has the effect
of introducing the edge $OE_{j+1}$ and replaces the two tetrahedra
by the three tetrahedra $ONE_jE_{j+1}$, $OSE_jE_{j+1}$ and $ONSE_{j+1}$.
This results in a triangulation of the 3-ball with $2p+1$ tetrahedra.
 Finally do an elementary move of order 2 on the 3 tetrahedra
$ONSE_n$, $ONSE_1$ and $NSE_1E_n$, replacing them by the two tetrahedra
$ONE_1E_n$ and $OSE_1E_n$.
\enddemo

\definition{Definition \propno} A singular manifold is a
complex with simplexes of dimension at most three, such that the
link of every edge is a circle and the link of every face is two points.
\enddefinition
A singular manifold is really a closed manifold with singularities.
The additional condition for a singular manifold to be a closed manifold
is that the link of every vertex is a 2-sphere. It follows from the
conditions that the link of a vertex in a singular manifold is a surface.
A singular manifold is a closed manifold if and only if its Euler number
is zero. The class of oriented singular manifolds can be characterised as those
complexes which are obtained from gluing the disjoint union of a number of
oriented tetrahedra by identifying faces pairwise by orientation-reversing
maps.

\proclaim{Theorem \propno \name\moves} Two triangulated singular manifolds
are piecewise-linearly homeomorphic if and only if they are related by a
finite sequence of elementary moves.\endproclaim

\demo{Proof} It is clear that if two singular manifolds are related by a
finite sequence of elementary moves then they are equivalent. It remains
to show the converse, that if two singular manifolds are related by a
finite sequence of stellar moves then they are related by a finite sequence
of elementary moves. It is sufficient to show that each stellar move can be
obtained as a finite sequence of elementary moves.

A stellar subdivision of order 3 is already an elementary move.
A stellar subdivision of order 2 and any stellar subdivision of order 1 are
done with the preceding lemmas.  \enddemo

The generalisation of theorem \moves\ from 3-manifolds to
arbitrary $n$-manifolds is the following and is proved in
\Lcitemark Pachner\Citebreak 1991\Rcitemark \Rspace{}. This theorem would be
the natural starting
point for constructing piecewise-linear field theories in
higher dimensions.
\proclaim{Theorem \propno} Two closed piecewise-linear $n$-manifolds
are equivalent if and only if they are related by a finite sequence
of elementary moves.\endproclaim

\head\headno Invariants of manifolds \endhead

The following is the main theorem in this paper.
\proclaim{Theorem \propno \name\plft} A finite semisimple spherical category
of non-zero dimension determines an invariant of oriented singular 3-manifolds.
\endproclaim

Since closed (3-)manifolds are examples of singular 3-manifolds, this
determines an invariant of closed manifolds. Throughout this section, the
proof refers to closed manifolds, which, as is the general convention
in this paper, are taken to be oriented.
However every statement is also true for oriented singular manifolds.

Let $M$ be a closed simplicial manifold, $J$ be a choice of one simple object
from each isomorphism class, and $K$ the dimension of the spherical category.

The notation in this section is as follows: for a simplicial manifold $M$,
the edge set is denoted $E$. Thus $l\colon E\to I$ is a labelling, and the
labelled manifold is the pair $(M,l)$. Let $v$ be the number of vertices
of $M$.

Define the state sum invariant of $M$ by a summation over the set of all
labellings by elements of $J$.
$$C(M)=K^{-v}
\sum_{l\colon E\to J}Z(M,l)\prod_{e\in E}\dim_q(l(e)).$$

\demo{Proof of theorem \plft} The rest of this section is a proof that
$C(M)$ is the manifold invariant. That is, that any simplicial manifold
$M$ which triangulates a given piecewise-linear manifold $\Cal M$ determines
the same invariant.

Let $M$ be a simplicial complex that triangulates $\Cal M$.
First,  $C(M)$ does not depend on the choice of
simple objects $J$ due to proposition \naturality.
Next it is necessary to show that $C(M)$ does not
depend on the choice of simplicial structure for the complex which
triangulates $\Cal M$, and finally that it does not depend
on the choice of triangulation.

Let $M_1$ and $M_2$ be two different choices of simplicial structure with
the same underlying complex and the same orientation. Then the identity map
of complexes is a combinatorial isomorphism of simplicial manifolds.
By theorem \symmetry, the state sum $C(M_1)$ is equal to a
state sum over the set of labellings of $M_2$ which are compatible with
a labelling $E\to J$ of $M_1$. This is not the state sum $C(M_2)$,
because the labelling of an edge in the complex runs over either
the set $J$ or the set $\hat J$=$\{\hat a\mid a\in J\}$. However it is equal to
$C(M_2)$ because $\hat a$ is a simple object if $a$ is, and $\hat J$
also contains one element of each isomorphism class of simple objects.
The equality follows from proposition \naturality. This shows that $C(M)$
does not depend on the simplicial structure. Now it remains to consider the
triangulation.

If $L$ is a subcomplex of a complex $M$, and $L$ has a simplicial structure
determined by a total order of the vertices of $L$, then this can be extended
to a simplicial structure of $M$, by extending the total order. If a complex
$N$ is obtained from the complex $M$ by an elementary move, so that
$M=X\cup\sigma^k*\partial\sigma^{3-k}$ and
$N=X\cup\partial\sigma^k*\sigma^{3-k}$, then a choice of standard simplicial
structure for $\partial\sigma^{4}$  can be extended to
$X\cup \partial\sigma^{4}$, which contains $M$ and $N$ as subcomplexes.
Such a choice of simplicial structure for $\partial\sigma^{4}$ is just the
identification of $\sigma^{4}$ as the boundary of the standard $4$-simplex,
$(01234)$.

Let $(ij)\mapsto e_{ij}$, for $0\le i<j\le 4$ be a labelling of the
standard 4-simplex $(01234)$. This determines  partition functions
for each tetrahedron in the boundary,
$$
Z(\pm(1234))=\left\{\matrix e_{12}&e_{13}&e_{23}\\e_{34}&e_{24}&e_{14}
\endmatrix\right\}_\pm \ \
Z(\pm(0234))=\left\{\matrix e_{02}&e_{03}&e_{23}\\e_{34}&e_{24}&e_{04}
\endmatrix\right\}_\pm$$$$
Z(\pm(0134))=\left\{\matrix e_{01}&e_{03}&e_{13}\\e_{34}&e_{14}&e_{04}
\endmatrix\right\}_\pm \ \
Z(\pm(0124))=\left\{\matrix e_{01}&e_{02}&e_{12}\\e_{24}&e_{14}&e_{04}
\endmatrix\right\}_\pm$$$$
Z(\pm(0123))=\left\{\matrix e_{01}&e_{02}&e_{12}\\e_{23}&e_{13}&e_{03}
\endmatrix\right\}_\pm.
$$

The invariance of $C(M)$ under elementary moves follows from the
next proposition.
\enddemo

Let $P$ be the map $x\otimes y\mapsto y\otimes x$.
\proclaim{Proposition \propno\name\identities}
\noindent {\bf (Orthogonality.)} The map
$$\dim_q(e_{02})\sum_{e_{13}\in J}Z(0123)Z(-0123)\dim_q(e_{13})$$
is equal to the identity map on
$H(e_{23},e_{03},e_{02})
\otimes H(e_{12},e_{02},e_{01})$.

\noindent {\bf (Biedenharn-Elliot.)} The equality
$$\multline \bigl(Z(0234)\otimes 1\bigr)\bigl(1\otimes Z(0124)\bigr)\\
=\sum_{e_{13}\in J}\dim_q(e_{13})\bigl(1\otimes Z(0123)\bigr)
\bigl(P\otimes 1\bigr)
\bigl(1\otimes Z(0134)\bigr)
\bigl(P\otimes 1\bigr)
\bigl(Z(1234)\otimes 1\bigr)\endmultline$$
holds.

These equalities hold for all choices of labels $\{e_{ij}\}$ not
explicitly summed over.
\endproclaim
The proof of these will be given below, after completing the proof of theorem
\plft. Theorem \plft\ follows once it has been established that $C(M)$
is invariant under elementary moves of order 2 and 3. The elementary moves
of order 1 and 0 are the inverses of these moves.

The simplicial invariant
can be decomposed as $Z(M,l_M)=\tr(Z(X),Z(D^1))$ and $Z(N,l_N)
=\tr(Z(X),Z(D^2))$, where $D^1$ and $D^2$ are the simplicial disks in
the elementary moves, $D^1\cup D^2=\partial (01234)$, $M=X\cup D^1$,
$N=X\cup D^2$, and $X\cup (01234)$ is labelled with restriction $l_M$ to $M$
and $l_N$ to $N$. The linear map $Z(X)$ is defined to be the partial trace
over the state spaces of all triangles not in the boundary of $X$ of the
tensor product of the partition
functions for each oriented labelled tetrahedron in $X$. The linear maps
$Z(D^1)$ and $Z(D^2)$ are defined likewise.

The invariance of $C(M)$
under elementary moves follows by establishing that
$$ K^{-v^1}\sum_l \left( Z(D^1)\prod_e\left(\dim_q(l(e))\right)\right)=
K^{-v^2}\sum_l \left( Z(D^2)\prod_e\left(\dim_q(l(e))\right)\right).$$
In this formula, $v^1$, $v^2$ are the number of vertices internal to
$D^1$, $D^2$ (i.e., not on the boundary); the product is over edges internal
to $D^1$ or $D^2$, and the summation is over labellings which are fixed
on $\partial D^1=\partial D^2$ but range over all values in $J$ for all
edges internal to $D^1$ or $D^2$.

For the elementary move of order 2, there are no internal vertices and the
equality is the Biedenharn-Elliot identity of proposition \identities.

For the elementary move of order 3, the required identity is

\proclaim{Lemma \propno}
$$\multline
Z(0234)=K^{-1}\sum_{e_{01}, e_{12},\atop e_{13}, e_{14}\in J}
\biggl(\tr_3\bigl(
\bigl(1\otimes Z(0123)\bigr)
\bigl(P\otimes 1\bigr)
\bigl(1\otimes Z(0134)\bigr)
\bigl(P\otimes 1\bigr)
 \\
\bigl(Z(1234)\otimes 1\bigr)
\bigl(1\otimes Z(-0124)\bigr)\bigr)
\prod_{n=0,2,3,4}\dim_q(e_{1n})\biggr).
\endmultline$$
in which $\tr_3$ is the partial trace over the third factor:
$$(\alpha\otimes\beta\otimes\gamma)\mapsto\alpha\otimes\beta \ \tr(\gamma).$$
\endproclaim

\demo{Proof}  Follow the linear maps on each side of the Biedenharn-Elliot
relation with the linear map
$$K^{-1}\dim_q\left(e_{01}\right)
\dim_q\left(e_{12}\right)\dim_q\left(e_{14}\right)
\left(1\otimes Z(-0124)\right),$$
take the partial trace on the third factor and sum over $e_{01}\in J$,
$e_{12}\in J$, and $e_{14}\in J$. The right hand side of the
Biedenharn-Elliot identity becomes the right hand side of the equation
in the statement of the lemma. The left hand side becomes
$$\multline K^{-1}\sum_{e_{01},e_{12}, e_{14}\in J}
\tr_3\bigl(\bigl(Z(0234)\otimes 1\bigr)\bigl(1\otimes Z(0124)\bigr)
\bigl(1\otimes Z(-0124)\bigr)\bigr)
\dim_q\left(e_{01}\right) \\
\dim_q\left(e_{12}\right)\dim_q\left(e_{14}\right).\endmultline$$
Using the orthogonality relation of proposition \identities, this is
equal to
$${K^{-1}\over\dim_q(e_{02})}\sum_{e_{01}, e_{12}\in J }
\tr_3\bigl(\bigl(Z(0234)\otimes 1\bigr)\bigr)
\dim_q\left(e_{01}\right)
\dim_q\left(e_{12}\right).$$
Also,
$$\tr_3\bigl(\bigl(Z(0234)\otimes 1\bigr)\bigr)=Z(0234) \dim
\Hom(e_{02},e_{12}\otimes e_{01}),$$
using the ordinary vector space dimension $\dim$. From the symmetry conditions
and lemma \dimensions, it follows that
$$\multline\sum_{e_{01}}\dim_q(e_{01})\dim\Hom(e_{02},e_{12}\otimes e_{01})=
\sum_{e_{01}}\dim_q(e_{01})\dim\Hom(e_{01},\hat e_{12}\otimes e_{02})\\
=\dim_q(e_{12})\dim_q(e_{02}).\endmultline$$
Thus the left hand side of the relation is equal to
$$K^{-1}Z(0234)\sum_{e_{12}\in J}\left(\dim_q\left(e_{12}\right)\right)^2
=Z(0234).$$
Thus the lemma is proved.\enddemo

To complete the proof of theorem \plft, it remains to prove proposition
\identities. The first step is

\proclaim {Lemma \propno{} (Crossing)\name\crossing}
The following diagram is commutative
$$\CD
{\bigoplus\atop{e_{02}\in J}} H(e_{23},e_{03},e_{02})
\otimes H(e_{12},e_{02},e_{01})
@>\Phi>>{\bigoplus\atop{e_{13}\in J}} H(e_{23},e_{13},e_{12} )
\otimes H(e_{13},e_{03},e_{01})\\
@V\alpha\otimes\beta\mapsto\alpha(1\otimes\beta)VV
@V\alpha\otimes\beta\mapsto\beta(\alpha\otimes 1)VV\\
\Hom(e_{03},e_{23}\otimes(e_{12}\otimes e_{01})) @=
\Hom(e_{03},(e_{23}\otimes e_{12})\otimes e_{01})
\endCD$$
where the linear map
$$\Phi=\bigoplus_{e_{02}\in J\atop e_{13}\in J} \dim_q(e_{13})
\left\{\matrix e_{01}&e_{02}&e_{12}\\e_{23}&e_{13}&e_{03}
\endmatrix\right\}_+.$$
Also, the  linear map
$$\Psi=\bigoplus_{e_{02}\in J\atop e_{13}\in J} \dim_q(e_{02})
\left\{\matrix e_{01}&e_{02}&e_{12}\\e_{23}&e_{13}&e_{03}
\endmatrix\right\}_-$$
is the inverse of $\Phi$. \endproclaim

\demo{Proof} The semisimple condition implies that the vertical arrows
are isomorphisms. Consider an element $\alpha\otimes\beta$ in the top
left-hand space. Taking the trace of the two images of this element
in the bottom right-hand space with $(\gamma\otimes 1)\delta$ for arbitrary
$\gamma\in\Hom(e_{23}\otimes e_{12},e_{13})$ and
$\delta\in\Hom(e_{13}\otimes e_{01},e_{03})$ yields two elements of $\ring$
which are equal. This shows that the diagram commutes.
Replacing the map $\Phi$ with the inverse of $\Psi$ also yields a
commutative diagram, by a similar argument. Combining these two
diagrams shows that $\Phi=\Psi^{-1}$.
\enddemo

\demo{Proof of proposition \identities} The proof of the orthogonality
relation is a direct consequence of lemma \crossing.
The Biedenharn-Elliot identity follows from the fact that the following
diagram commutes. In this diagram the shorthand notation
$(012)$ is used for the state space of this triangle,
$\Hom(e_{02},e_{12}\otimes e_{01})$.
%
%
\comment
$$\diagram
{\bigoplus\atop{e_{02},e_{03}\in J}}
(034)\otimes (023)\otimes (012)
\rrto^{1\otimes Z(0123)}
\ddto_{Z(0234)\otimes 1} & &
{\bigoplus\atop{e_{03},e_{13}\in J}}
(034)\otimes (123)\otimes (013)
 \dto^{(P\otimes 1)(1\otimes Z(0134))}\\
& & {\bigoplus\atop{e_{13},e_{14}\in J}}
 (123)\otimes (134) \otimes (014)
\dto^{PZ(1234)\otimes 1}\\
{\bigoplus\atop{e_{02},e_{24}\in J}}
(234)\otimes(024)\otimes (012)
 \rrto^{1\otimes Z(0124)} & &
{\bigoplus\atop{e_{14},e_{24}\in J}}
 (234)\otimes(124) \otimes(014)
\enddiagram$$
\endcomment
%
%
$$\CD
{\bigoplus\atop{e_{02},e_{03}\in J}}
(034)\otimes (023)\otimes (012)
@>{1\otimes Z(0123)}>>
{\bigoplus\atop{e_{03},e_{13}\in J}}
(034)\otimes (123)\otimes (013)   \\
@V{Z(0234)\otimes 1}VV
@VV{(P\otimes 1)(1\otimes Z(0134))}V\\
@.{\bigoplus\atop{e_{13},e_{14}\in J}}
 (123)\otimes (134) \otimes (014)\\
@VVV @VV{PZ(1234)\otimes 1}V\\
{\bigoplus\atop{e_{02},e_{24}\in J}}
(234)\otimes(024)\otimes (012)
 @>{1\otimes Z(0124)}>>
{\bigoplus\atop{e_{14},e_{24}\in J}}
 (234)\otimes(124) \otimes(014)
\endCD$$
%
%
Applying lemma \crossing{} five times shows that this diagram commutes.
\enddemo

\remark{Remarks } The idea of using a state sum model to construct
manifold invariants is due to\Lspace \Lcitemark Turaev\Nameand Viro\Citebreak
1992\Rcitemark \Rspace{}. where an invariant is constructed
from the $6j$-symbols of $U_q\SL_2$,
for $q$ a root of unity.
 Every state space
$H(a,b,c)$ is 0 or $\C$ and every map $\theta(\sigma)\colon
V{\sigma(a,b,c)}\to V(a,b,c)$ is the identity. Hence the
partition function of a labelled tetrahedron is simply a number, and these
numbers are equal under permutations of the labels.
These numbers are the quantum analogues of the $6j$-symbols.
The identities of proposition \identities\ are the quantum analogues
of the well-known Biedenharn-Elliot and
orthogonality identities, as proved in\Lspace \Lcitemark Reshetikhin\Nameand
Kirillov\Citebreak 1988\Rcitemark \Rspace{}.

This example satisifies some extra conditions, which entail firstly
that the invariant is defined for unoriented manifolds, and secondly that
there is a topological quantum field theory associated to the invariant.

The first condition is that each space $H(a,b,c)$ is an inner product space
and that the partition function of $-A$ is the adjoint of the partition
function of $A$ with respect to these inner products. This condition
implies that the invariant is defined for unoriented manifolds.

The second condition is that each self-dual simple object in the category is
orthogonal and not symplectic. If $a$ is a self-dual simple object, then
there is an isomorphism $\phi\colon a\to\hat a$. This object is called
orthogonal if $\phi=\hat\phi$, or symplectic if $\phi=-\hat\phi$. These
are the only possibilities as $\spacehat$ is an involution. This
classification does not depend on the choice of isomorphism $\phi$ as
$\Hom(a,\hat a)\cong \ring$ and $\spacehat$ is linear.

This condition is necessary for the construction of a topological field
theory by a construction similar to that of\Lspace \Lcitemark Turaev\Nameand
Viro\Citebreak 1992\Rcitemark \Rspace{}, because it is necessary
for the isomorphisms of naturality (definition \naturaliso) and symmetry
(definition \action) on the state space of a
triangle to commute. The condition is a sufficient condition because it is
a coherence condition which
allows the construction of a strict spherical category in which $\hat a=a$
for all self-dual objects.

For the example of $U_q\SL_2$, there are exactly two choices of the
element $w$ which make this Hopf algebra spherical. The topological field
theory of\Lspace \Lcitemark Turaev\Nameand Viro\Citebreak 1992\Rcitemark
\Rspace{} is constructed with the element for which every simple
module is orthogonal. At the value $q=1$, this element $w$ takes the
value $1$ on the even (integer spin) representations and $-1$ on the
odd (odd half-integer spin) representations.
\endremark

\head\headno Spherical Hopf algebras \endhead
\definition{Definition \propno \name\sphericalHopf} A spherical Hopf algebra
over a field $k$ consists of a finite dimensional vector space $A$
together with the following data \roster
\item a multiplication $\mu$
\item a unit $\eta\colon \ring\to A$
\item a comultiplication $\Delta\colon A\to A\otimes A$
\item a counit $\epsilon\colon A\to \ring$
\item an antipodal map $\gamma\colon A\to A$
\item an element $w\in A$
\endroster\enddefinition

The data $(A,\mu ,\eta ,\Delta ,\epsilon ,\gamma )$ is required to define a
Hopf algebra. The conditions on the element $w$ are the following:
\roster
\item $\gamma^2(a)=waw^{-1}$ for all $a\in A$.
\item $\Delta (w)=w\otimes w$.
\item $\tr (\theta w)=\tr (\theta w^{-1})$ for all left $A$-modules $V$
and all $\theta\in\End_A(V)$.
\endroster
It follows from the condition $\Delta(w)=w\otimes w$ that $\gamma (w)=w^{-1}$
and that $\epsilon(w)=1$. Such elements are called group-like.
\example{Example \propno} Examples of Hopf algebras which are spherical are:
\roster
\item Any involutory Hopf algebra is spherical. The element $w$ can be taken
to be 1.
\item Any ribbon Hopf algebra, as defined in\Lspace \Lcitemark
Reshetikhin\Nameand Turaev\Citebreak 1990\Rcitemark \Rspace{},
is spherical. The element $w$ can be taken to be $uv^{-1}$ where the
element $u$ is determined by the quasi-triangular structure and the
element $v$ is the ribbon element.
\endroster\endexample

\remark{Remark \propno} A Hopf algebra with an element $w$ that satisfies
the first two conditions of definition \sphericalHopf\  is spherical if,
either $w^2=1$, or all modules are isomorphic to their dual.\endremark

\remark{Remark \propno} If $A$ is a Hopf algebra there may exist more
than one element $w$ such that $(A,w)$ is a spherical Hopf algebra.
However, if $w_1$ is one such element then $w_2=gw_1$ is another such
element if and only if $g$ satisfies the conditions:
\roster
\item $g$ is central
\item $g$ is group-like
\item $g$ is an involution
\endroster
\endremark

\example{Example \propno} This is an example of a finite dimensional
Hopf algebra which satisfies all the conditions for a spherical Hopf
algebra except that the left and right traces are distinct. This example
is the quantised enveloping algebra of the Borel subalgebra of $\SL_2(\C)$.

Let $s$ be a primitive $2r$-th root of unity
with $r>1$. Let $B$ be the unital algebra generated by
elements $X$ and $K$ subject to the defining relations
$$\align
KX&=sXK\\
K^{4r}&=1\\
X^r&=0
\endalign$$

Then $B$ is a finite dimensional algebra and also has a Hopf algebra
structure defined by:
\roster
\item The coproduct, $\Delta$, is defined by
$\Delta (K)= K\otimes K$ and $\Delta (X)= X\otimes K + K^{-1}\otimes X$
\item The augmentation, $\epsilon$, is defined by
$\epsilon (K)=1$ and $\epsilon (X)=0$
\item The antipode, $\gamma$, is defined by
$\gamma (K)= K^{-1}$ and $\gamma (X)= -sX$
\endroster

The element $w=K^2$ satisfies the conditions
$$\align
\Delta (w)&=w\otimes w\\
\epsilon (w)&=1\\
\gamma (w)&=w^{-1}\\
\gamma^2 (b)&=wbw^{-1}\quad\text{for all $b\in B$}
\endalign$$

The trace condition is not satisfied since $B$
has $4r$ one dimensional representations with $X=0$ and $K$ a $4r$-th
root of unity and it is clear that the trace condition is not satisfied
in these representations.
\endexample

\example{Example \propno} Not all spherical Hopf algebras are modular
Hopf algebras in the sense of\Lspace \Lcitemark Reshetikhin\Nameand
Turaev\Citebreak 1991\Rcitemark \Rspace{}. For example,
the group algebra of a finite group over a field of characteristic 0,
or more generally, any cocommutative, involutory, semisimple Hopf algebra,
is spherical and not modular.

Also, not all spherical Hopf algebras are ribbon Hopf algebras. The dual of
the group algebra of a non-commutative finite group can be made a
spherical Hopf algebra. However, the representation ring
is non-commutative, and so the Hopf algebra cannot be quasi-triangular.
\endexample

The next proposition gives the application of spherical Hopf algebras to
state sum models.

\proclaim{Proposition \propno} The category of representations of a
spherical Hopf algebra, $A$, is equivalent to a canonical strict
spherical category.
\endproclaim

In this category, the objects are lists of left $A$-modules and the
morphisms $A$-linear maps of the modules formed by tensor product over the
list. The trace $\tr_L$ of an endomorphism $\theta$ of a left A-module
is the matrix trace of $\theta w$.
The full construction is given in\Lspace \Lcitemark Barrett\Nameand
Westbury\Citebreak 1993\Rcitemark \Rspace{}.

Some extra conditions are required to give invariants of manifolds. The
first way this can be achieved uses the same data as the construction of
invariants given in\Lspace \Lcitemark Kuperberg\Citebreak 1991\Rcitemark
\Rspace{}

\proclaim{Proposition \propno\name\invol} Let $A$ be a finite dimensional
involutory Hopf algebra over an algebraically closed field of characteristic 0.
Then $A$ determines a 3-manifold invariant.
\endproclaim
\demo{Proof} The Hopf algebra $A$ gives a spherical Hopf algebra by taking
$w=1$.
In this case the quantum trace is just the matrix trace. The algebra $A$ is
semisimple by\Lspace \Lcitemark Larson\Nameand Radford\Citebreak 1988\Rcitemark
\Rspace{}. Hence the category of finite dimensional left
$A$-modules is a spherical category which satisfies the hypotheses of theorem
\plft .
\enddemo

In general, the construction of the non-degenerate quotient is another way
of attaining the semisimple condition.
The following proposition is proved in\Lspace \Lcitemark Barrett\Nameand
Westbury\Citebreak 1993\Rcitemark \Rspace{}.

\proclaim{Proposition \propno} Let $A$ be a spherical Hopf algebra over an
algebraically closed field. Then the non-degenerate quotient of the spherical
category of finitely generated left $A$-modules is semisimple.
\endproclaim

A corollary to the proof of this proposition is that the nondegenerate quotient
of any spherical subcategory of the category of left $A$-modules which is
closed
under taking direct summands is a semisimple spherical category. Thus, in order
to construct a manifold invariant, it is sufficient to construct such spherical
subcategories which are finite and have non-zero dimension.

Let $A$ be a finite dimensional spherical Hopf algebra. If $A$ is semisimple as
an algebra then it is clear that the non-degenerate quotient of the category of
left $A$-modules satisfies all the hypotheses of theorem \plft\ except possibly
the condition that $K=0$. In the following discussion we assume that $A$ is not
semisimple. In this case the set of isomorphism classes of simple objects in
the non-degenerate quotient may well be infinite. This may come about since
each $A$-module $V$ that satisfies $\End_A(V)\cong\ring$ and which has non-zero
quantum dimension gives a simple object in the quotient. The condition that
$\End_A(V)\cong\ring$ is much weaker than the condition that $V$ is
irreducible. Although it is possible for inequivalent modules to give
equivalent
simple objects in the quotient, there is no apparent reason for the set of
simple objects in the quotient to be finite, in general.

Hence in order to construct a manifold invariant from a spherical Hopf
algebra $A$ which is not semisimple it is necessary to find a proper spherical
subcategory of the category of left $A$-modules such that the non-degenerate
quotient is finite and has non-zero dimension. A subcategory of a
spherical category is a spherical subcategory if and only if it is closed under
addition, tensor product and taking duals.

The category of left $A$-modules has a proper spherical subcategory namely the
category of projective left $A$-modules. This category is spherical since it
is closed under tensor product and is closed under taking duals by\Lspace
\Lcitemark Larson\Nameand Sweedler\Citebreak 1969\Rcitemark \Rspace{}.
However the next proposition is a negative result which shows that this
spherical subcategory cannot be used to construct manifold invariants.

\proclaim{Proposition \propno} Let $A$ be a finite dimensional pivotal
Hopf algebra which is not semisimple. Then every projective $A$-module has
zero quantum dimension.
\endproclaim

\demo{Proof} If $A$ is any Hopf algebra then the linear functional
$$a\mapsto \tr (x\mapsto \gamma^{-2}(xa))$$
is a right co-integral.
If $A$ is not a semisimple algebra this is identically zero\Lspace \Lcitemark
Larson\Nameand Radford\Citebreak 1988\LIcitemark{}, proposition 2.4\RIcitemark
\Rcitemark \Rspace{}; and so if
$A$ satisfies the hypotheses of the proposition then, for all $a\in A$,
$\tr (x\mapsto w^{-1}xaw)=0$.
Now choose a primitive idempotent $\pi$ and put $a=\pi\omega^{-1}$. This gives
$$0=\tr (x\mapsto w^{-1}x\pi)=\dim_q(A\pi)$$
which shows that every indecomposable projective $A$-module has zero quantum
dimension.
\enddemo

Examples of spherical Hopf algebras which are not semisimple are quantised
enveloping algebras at roots of unity. The following theorem shows, following
\Lcitemark Andersen\Citebreak 1992\Rcitemark \Rspace{}, that each of these
examples gives a manifold invariant.
\proclaim{Theorem \propno }
Let $A$ be a quantised enveloping algebra of a finite dimensional semisimple
Lie
algebra at a root of unity. Assume that the order, $k$, of the quantum
parameter
$q$ is at least the Coxeter number of the Lie algebra; that this order is odd;
and that $k$ is not divisible by three if the factor $G_2$ occurs. Then the
category of tilting modules is spherical and the non-degenerate quotient
is finite with non-zero dimension, satisfying the hypotheses of theorem \plft .
\endproclaim

\demo{Proof} It follows from the definition that the category of tilting
modules
is closed under taking duals and direct summands. It follows from the deep
result of Lusztig on canonical bases that the category of tilting modules is
closed under tensor product. The isomorphism classes of indecomposable tilting
modules are indexed by the dominant weights and it is shown in
\Lcitemark Andersen\Citebreak 1992\Rcitemark \Rspace{}
that the quantum dimension is non-zero if and only if the dominant weight lies
in the interior of the fundamental alcove. In particular this shows that the
set of isomorphism classes of simple objects in the non-degenerate quotient
is finite. Each tilting module corresponding to a dominant weight in the
interior of the fundamental alcove is a Weyl module (as well as being
irreducible) and so the quantum dimension is given by the Kac formula. The
quantum dimension of the dual is given by substituting $q^{-1}$ for $q$ in this
formula. Since $\dim_q(V)=\dim_q(V^*)$ for any $V$ and $q$ has unit modulus it
follows that each quantum dimension is real. Hence the dimension $K$ is a
sum of positive real numbers. The condition that $k$ is greater than or equal
to the Coxeter number ensures that there is at least one dominant weight
in the interior of the fundamental alcove, and so $K$ is non-zero.
\enddemo

\head References \endhead

\message{REFERENCE LIST}

\bgroup\Resetstrings%
\def\Ecnt{0}\def\acnt{0}%
\def\Ftest{ }\def\Ftrail{0}\def\Fstr{Alexander\Citebreak 1930}%
\def\Atest{ }\def\Astr{J\Initper \Initgap W\Initper  Alexander}%
\def\Ttest{ }\def\Tstr{The combinatorial theory of complexes}%
\def\Jtest{ }\def\Jstr{Ann. of Math. (2)}%
\def\Dtest{ }\def\Dstr{1930}%
\def\Vtest{ }\def\Vstr{31}%
\def\Ptest{ }\def\Pstr{294--322}%
\Refformat\egroup%

\bgroup\Resetstrings%
\def\Ecnt{0}\def\acnt{0}%
\def\Ftest{ }\def\Ftrail{2}\def\Fstr{Andersen\Citebreak 1992}%
\def\Atest{ }\def\Astr{H\Initper \Initgap H\Initper  Andersen}%
\def\Ttest{ }\def\Tstr{Tensor products of quantized tilting modules}%
\def\Jtest{ }\def\Jstr{Comm. Math. Phys.}%
\def\Vtest{ }\def\Vstr{149}%
\def\Dtest{ }\def\Dstr{1992}%
\def\Ptest{ }\def\Pstr{149--159}%
\Refformat\egroup%

\bgroup\Resetstrings%
\def\Ecnt{0}\def\acnt{0}%
\def\Ftest{ }\def\Ftrail{ }\def\Fstr{Barrett\Nameand Westbury\Citebreak 1993}%
\def\Atest{ }\def\Astr{J\Initper \Initgap W\Initper  Barrett%
  \Aand B\Initper \Initgap W\Initper  Westbury }%
\def\Ttest{ }\def\Tstr{Spherical categories}%
\def\Itest{ }\def\Istr{University of Nottingham}%
\def\Otest{ }\def\Ostr{preprint, hep-th/9310164}%
\def\Dtest{ }\def\Dstr{1993}%
\Refformat\egroup%

\bgroup\Resetstrings%
\def\Ecnt{0}\def\acnt{0}%
\def\Ftest{ }\def\Ftrail{ }\def\Fstr{Barrett\Nameand Westbury\Citebreak 1994}%
\def\Atest{ }\def\Astr{J\Initper \Initgap W\Initper  Barrett%
  \Aand B\Initper \Initgap W\Initper  Westbury }%
\def\Ttest{ }\def\Tstr{The equality of 3-manifold invariants}%
\def\Itest{ }\def\Istr{University of Nottingham}%
\def\Otest{ }\def\Ostr{preprint, hep-th/9406019}%
\def\Dtest{ }\def\Dstr{1994}%
\Refformat\egroup%

\bgroup\Resetstrings%
\def\Ecnt{0}\def\acnt{0}%
\def\Ftest{ }\def\Ftrail{3}\def\Fstr{Durhuus\Namecomma Jakobsen\Nameandd
Nest\Citebreak 1993}%
\def\Atest{ }\def\Astr{B\Initper  Durhuus%
  \Acomma H\Initper \Initgap P\Initper  Jakobsen%
  \Aandd R\Initper  Nest}%
\def\Ttest{ }\def\Tstr{Topological quantum field theories from generalized
$6j$-symbols}%
\def\Jtest{ }\def\Jstr{Reviews in Math. Physics}%
\def\Vtest{ }\def\Vstr{5}%
\def\Dtest{ }\def\Dstr{1993}%
\def\Ptest{ }\def\Pstr{1--67}%
\Refformat\egroup%

\bgroup\Resetstrings%
\def\Ecnt{0}\def\acnt{0}%
\def\Ftest{ }\def\Ftrail{9}\def\Fstr{Freyd\Nameand Yetter\Citebreak 1989}%
\def\Atest{ }\def\Astr{P\Initper \Initgap J\Initper  Freyd%
  \Aand D\Initper \Initgap N\Initper  Yetter}%
\def\Ttest{ }\def\Tstr{Braided compact closed categories with applications to
low dimensional topology}%
\def\Jtest{ }\def\Jstr{Adv. in Math.}%
\def\Vtest{ }\def\Vstr{77}%
\def\Dtest{ }\def\Dstr{1989}%
\def\Ptest{ }\def\Pstr{156--182}%
\Refformat\egroup%

\bgroup\Resetstrings%
\def\Ecnt{0}\def\acnt{0}%
\def\Ftest{ }\def\Ftrail{2}\def\Fstr{Freyd\Nameand Yetter\Citebreak 1992}%
\def\Atest{ }\def\Astr{P\Initper  Freyd%
  \Aand D\Initper \Initgap N\Initper  Yetter}%
\def\Ttest{ }\def\Tstr{Coherence theorems via knot theory}%
\def\Jtest{ }\def\Jstr{J. Pure Appl. Algebra}%
\def\Vtest{ }\def\Vstr{78}%
\def\Dtest{ }\def\Dstr{1992}%
\def\Ptest{ }\def\Pstr{49--76}%
\Refformat\egroup%

\bgroup\Resetstrings%
\def\Ecnt{0}\def\acnt{0}%
\def\Ftest{ }\def\Ftrail{0}\def\Fstr{Glaser\Citebreak 1970}%
\def\Atest{ }\def\Astr{L\Initper \Initgap C\Initper  Glaser}%
\def\Btest{ }\def\Bstr{Geometrical Combinatorial Topology I}%
\def\Dtest{ }\def\Dstr{1970}%
\def\Stest{ }\def\Sstr{Van Nostrand Reinhold Mathematical Studies}%
\def\Vtest{ }\def\Vstr{27}%
\Refformat\egroup%

\bgroup\Resetstrings%
\def\Ecnt{0}\def\acnt{0}%
\def\Ftest{ }\def\Ftrail{1}\def\Fstr{Joyal\Nameand Street\Citebreak 1991}%
\def\Atest{ }\def\Astr{A\Initper  Joyal%
  \Aand R\Initper  Street}%
\def\Ttest{ }\def\Tstr{The geometry of tensor calculus,I}%
\def\Jtest{ }\def\Jstr{Adv. in Math.}%
\def\Vtest{ }\def\Vstr{88}%
\def\Dtest{ }\def\Dstr{1991}%
\def\Ptest{ }\def\Pstr{55--112}%
\Refformat\egroup%

\bgroup\Resetstrings%
\def\Ecnt{0}\def\acnt{0}%
\def\Ftest{ }\def\Ftrail{0}\def\Fstr{Kelly\Nameand Laplaza\Citebreak 1980}%
\def\Atest{ }\def\Astr{G\Initper \Initgap M\Initper  Kelly%
  \Aand M\Initper \Initgap I\Initper  Laplaza}%
\def\Ttest{ }\def\Tstr{Coherence for compact closed categories}%
\def\Jtest{ }\def\Jstr{J. Pure Appl. Algebra}%
\def\Vtest{ }\def\Vstr{19}%
\def\Dtest{ }\def\Dstr{1980}%
\def\Ptest{ }\def\Pstr{193--213}%
\Refformat\egroup%

\bgroup\Resetstrings%
\def\Ecnt{0}\def\acnt{0}%
\def\Ftest{ }\def\Ftrail{1}\def\Fstr{Kuperberg\Citebreak 1991}%
\def\Atest{ }\def\Astr{G\Initper  Kuperberg}%
\def\Ttest{ }\def\Tstr{Involutory Hopf algebras and 3-manifold invariants}%
\def\Jtest{ }\def\Jstr{International Journal of Mathematics}%
\def\Vtest{ }\def\Vstr{2}%
\def\Dtest{ }\def\Dstr{1991}%
\def\Ptest{ }\def\Pstr{41--66}%
\Refformat\egroup%

\bgroup\Resetstrings%
\def\Ecnt{0}\def\acnt{0}%
\def\Ftest{ }\def\Ftrail{8}\def\Fstr{Larson\Nameand Radford\Citebreak 1988}%
\def\Atest{ }\def\Astr{R\Initper \Initgap G\Initper  Larson%
  \Aand D\Initper \Initgap E\Initper  Radford}%
\def\Ttest{ }\def\Tstr{Finite dimensional cosemisimple Hopf algebras in
characteristic 0 are semisimple}%
\def\Jtest{ }\def\Jstr{J. Algebra}%
\def\Dtest{ }\def\Dstr{1988}%
\def\Vtest{ }\def\Vstr{117}%
\def\Ptest{ }\def\Pstr{267--289}%
\Refformat\egroup%

\bgroup\Resetstrings%
\def\Ecnt{0}\def\acnt{0}%
\def\Ftest{ }\def\Ftrail{9}\def\Fstr{Larson\Nameand Sweedler\Citebreak 1969}%
\def\Atest{ }\def\Astr{R\Initper \Initgap G\Initper  Larson%
  \Aand M\Initper \Initgap E\Initper  Sweedler}%
\def\Ttest{ }\def\Tstr{An associative orthogonal bilinear form for Hopf
algebras}%
\def\Jtest{ }\def\Jstr{Amer. J. Math.}%
\def\Vtest{ }\def\Vstr{91}%
\def\Dtest{ }\def\Dstr{1969}%
\def\Ptest{ }\def\Pstr{75--94}%
\Refformat\egroup%

\bgroup\Resetstrings%
\def\Ecnt{0}\def\acnt{0}%
\def\Ftest{ }\def\Ftrail{3}\def\Fstr{Moussouris\Citebreak 1983}%
\def\Atest{ }\def\Astr{J\Initper \Initgap P\Initper  Moussouris}%
\def\Ttest{ }\def\Tstr{Quantum models of space-time based on coupling theory}%
\def\Jtest{ }\def\Jstr{D. Phil. Oxford}%
\def\Dtest{ }\def\Dstr{1983}%
\Refformat\egroup%

\bgroup\Resetstrings%
\def\Ecnt{0}\def\acnt{0}%
\def\Ftest{ }\def\Ftrail{1}\def\Fstr{Pachner\Citebreak 1991}%
\def\Atest{ }\def\Astr{U\Initper  Pachner}%
\def\Ttest{ }\def\Tstr{P.L. homeomorphic manifolds are equivalent by elementary
shellings}%
\def\Itest{ }\def\Istr{Academic Press}%
\def\Ctest{ }\def\Cstr{New York, NY}%
\def\Jtest{ }\def\Jstr{European Journal of Combinatorics}%
\def\Vtest{ }\def\Vstr{12}%
\def\Dtest{ }\def\Dstr{1991}%
\def\Ptest{ }\def\Pstr{129--145}%
\Refformat\egroup%

\bgroup\Resetstrings%
\def\Ecnt{0}\def\acnt{0}%
\def\Ftest{ }\def\Ftrail{8}\def\Fstr{Ponzano\Nameand Regge\Citebreak 1968}%
\def\Atest{ }\def\Astr{G\Initper  Ponzano%
  \Aand T\Initper  Regge}%
\def\Ttest{ }\def\Tstr{Semiclassical limit of Racah coefficients}%
\def\Btest{ }\def\Bstr{Spectroscopic and group theoretical methods in physics}%
\def\Dtest{ }\def\Dstr{1968}%
\def\Ptest{ }\def\Pstr{1--58}%
\def\Itest{ }\def\Istr{North-Holland}%
\def\Ctest{ }\def\Cstr{Amsterdam}%
\Refformat\egroup%

\bgroup\Resetstrings%
\def\Ecnt{0}\def\acnt{0}%
\def\Ftest{ }\def\Ftrail{1}\def\Fstr{Reshetikhin\Nameand Turaev\Citebreak
1991}%
\def\Atest{ }\def\Astr{N\Initper  Reshetikhin%
  \Aand V\Initper \Initgap G\Initper  Turaev}%
\def\Ttest{ }\def\Tstr{Invariants of 3-manifolds via link polynomials and
quantum groups}%
\def\Jtest{ }\def\Jstr{Invent. Math.}%
\def\Vtest{ }\def\Vstr{103}%
\def\Dtest{ }\def\Dstr{1991}%
\def\Ptest{ }\def\Pstr{547--597}%
\Refformat\egroup%

\bgroup\Resetstrings%
\def\Ecnt{0}\def\acnt{0}%
\def\Ftest{ }\def\Ftrail{8}\def\Fstr{Reshetikhin\Nameand Kirillov\Citebreak
1988}%
\def\Atest{ }\def\Astr{N\Initper \Initgap Y\Initper  Reshetikhin%
  \Aand A\Initper \Initgap N\Initper  Kirillov}%
\def\Ttest{ }\def\Tstr{Representations of the algebra $U_q(sl(2))$,
$q$-orthogonal polynomials and invariants of links}%
\def\Jtest{ }\def\Jstr{LOMI preprint}%
\def\Dtest{ }\def\Dstr{1988}%
\def\Ntest{ }\def\Nstr{E-9-88}%
\Refformat\egroup%

\bgroup\Resetstrings%
\def\Ecnt{0}\def\acnt{0}%
\def\Ftest{ }\def\Ftrail{0}\def\Fstr{Reshetikhin\Nameand Turaev\Citebreak
1990}%
\def\Atest{ }\def\Astr{N\Initper \Initgap Y\Initper  Reshetikhin%
  \Aand V\Initper \Initgap G\Initper  Turaev}%
\def\Ttest{ }\def\Tstr{Ribbon graphs and their invariants derived from quantum
groups}%
\def\Jtest{ }\def\Jstr{Comm. Math. Phys.}%
\def\Vtest{ }\def\Vstr{127}%
\def\Dtest{ }\def\Dstr{1990}%
\def\Ptest{ }\def\Pstr{1--26}%
\Refformat\egroup%

\bgroup\Resetstrings%
\def\Ecnt{0}\def\acnt{0}%
\def\Ftest{ }\def\Ftrail{3}\def\Fstr{Roberts\Citebreak 1993}%
\def\Atest{ }\def\Astr{J\Initper  Roberts}%
\def\Ttest{ }\def\Tstr{Skein theory and Turaev-Viro invariants}%
\def\Dtest{ }\def\Dstr{1993}%
\def\Otest{ }\def\Ostr{preprint, University of Cambridge}%
\Refformat\egroup%

\bgroup\Resetstrings%
\def\Ecnt{0}\def\acnt{0}%
\def\Ftest{ }\def\Ftrail{2}\def\Fstr{Rourke\Nameand Sanderson\Citebreak 1982}%
\def\Atest{ }\def\Astr{C\Initper \Initgap P\Initper  Rourke%
  \Aand B\Initper \Initgap J\Initper  Sanderson}%
\def\Ttest{ }\def\Tstr{Introduction to piece-wise linear topology}%
\def\Dtest{ }\def\Dstr{1982}%
\def\Itest{ }\def\Istr{Springer-Verlag}%
\def\Ctest{ }\def\Cstr{New York--Heidelberg--Berlin}%
\Refformat\egroup%

\bgroup\Resetstrings%
\def\Ecnt{0}\def\acnt{0}%
\def\Ftest{ }\def\Ftrail{2}\def\Fstr{Turaev\Nameand Viro\Citebreak 1992}%
\def\Atest{ }\def\Astr{V\Initper \Initgap G\Initper  Turaev%
  \Aand O\Initper \Initgap Y\Initper  Viro}%
\def\Ttest{ }\def\Tstr{State sum invariants of 3-manifolds and quantum
$6j$-symbols}%
\def\Jtest{ }\def\Jstr{Topology}%
\def\Vtest{ }\def\Vstr{31}%
\def\Ptest{ }\def\Pstr{865--902}%
\def\Dtest{ }\def\Dstr{1992}%
\Refformat\egroup%

\bgroup\Resetstrings%
\def\Ecnt{0}\def\acnt{0}%
\def\Ftest{ }\def\Ftrail{2}\def\Fstr{Turaev\Citebreak 1992}%
\def\Atest{ }\def\Astr{V\Initper  Turaev}%
\def\Ttest{ }\def\Tstr{Quantum invariants of 3-manifold and a glimpse of shadow
topology}%
\def\Jtest{ }\def\Jstr{Lect. Notes in Math.}%
\def\Itest{ }\def\Istr{Springer-Verlag}%
\def\Ctest{ }\def\Cstr{New York--Heidelberg--Berlin}%
\def\Dtest{ }\def\Dstr{1992}%
\def\Vtest{ }\def\Vstr{1510}%
\def\Ptest{ }\def\Pstr{363--366}%
\Refformat\egroup%

\bgroup\Resetstrings%
\def\Ecnt{0}\def\acnt{0}%
\def\Ftest{ }\def\Ftrail{3}\def\Fstr{Turaev\Nameand Wenzl\Citebreak 1993}%
\def\Atest{ }\def\Astr{V\Initper  Turaev%
  \Aand H\Initper  Wenzl}%
\def\Ttest{ }\def\Tstr{Quantum invariants of 3-manifolds associated with
classical simple Lie algebras}%
\def\Jtest{ }\def\Jstr{International Journal of Mathematics}%
\def\Dtest{ }\def\Dstr{1993}%
\def\Vtest{ }\def\Vstr{4}%
\def\Ptest{ }\def\Pstr{323--358}%
\Refformat\egroup%

\bgroup\Resetstrings%
\def\Ecnt{0}\def\acnt{0}%
\def\Ftest{ }\def\Ftrail{2}\def\Fstr{Turaev\Citebreak 1992}%
\def\Atest{ }\def\Astr{V\Initper \Initgap G\Initper  Turaev}%
\def\Ttest{ }\def\Tstr{Modular categories and 3-manifold invariants}%
\def\Jtest{ }\def\Jstr{Internat. J. Modern Phys.}%
\def\Vtest{ }\def\Vstr{6}%
\def\Dtest{ }\def\Dstr{1992}%
\def\Ptest{ }\def\Pstr{1807--1824}%
\Refformat\egroup%

\bgroup\Resetstrings%
\def\Ecnt{0}\def\acnt{0}%
\def\Ftest{ }\def\Ftrail{0}\def\Fstr{Walker\Citebreak 1990}%
\def\Atest{ }\def\Astr{K\Initper  Walker}%
\def\Ttest{ }\def\Tstr{On Witten's 3-manifold invariants}%
\def\Dtest{ }\def\Dstr{1990}%
\Refformat\egroup%

\bgroup\Resetstrings%
\def\Ecnt{0}\def\acnt{0}%
\def\Ftest{ }\def\Ftrail{ }\def\Fstr{Yetter\Citebreak 1993}%
\def\Atest{ }\def\Astr{D\Initper \Initgap N\Initper  Yetter }%
\def\Ttest{ }\def\Tstr{State-sum invariants of 3-manifolds associated to
Artinian semisimple tortile categories}%
\def\Jtest{ }\def\Jstr{Topology Appl.}%
\def\Otest{ }\def\Ostr{(to appear)}%
\def\Dtest{ }\def\Dstr{1993}%
\Refformat\egroup%

\tracingstats=1
\enddocument